\documentclass[aps,preprint]{revtex4}%
\usepackage{amsfonts}
\usepackage{amsmath}
\usepackage{amssymb}
\usepackage{graphicx}%
\begin{document}
\preprint{ }
\title{Lattice calculations for $A=3,4,6,12$ nuclei \\using chiral effective field theory}
\author{{Evgeny Epelbaum$^{a}$, Hermann~Krebs$^{a}$, Dean~Lee$^{b,c}$,
Ulf-G.~Mei{\ss }ner$^{c,d,e}$} \linebreak}
\affiliation{$^{a}$Institut f\"{u}r Theoretische Physik II, Ruhr-Universit\"{a}t Bochum,
D-44780 Bochum, Germany \linebreak$^{b}$Department of Physics, North Carolina
State University, Raleigh, NC 27695, USA \linebreak$^{c}$Helmholtz-Institut
f\"{u}r Strahlen- und Kernphysik (Theorie) and Bethe Center for Theoretical
Physics, Universit\"{a}t Bonn, D-53115 Bonn, Germany \linebreak$^{d}$Institut
f\"{u}r Kernphysik (IKP-3) and J\"{u}lich Center for Hadron Physics,
\ Forschungszentrum J\"{u}lich, D-52425 J\"{u}lich, Germany \linebreak$^{e}%
$Institute for Advanced Simulation (IAS-4), Forschungszentrum J\"{u}lich,
D-52425 J\"{u}lich, Germany\medskip\medskip}

\begin{abstract}
We present lattice calculations for the ground state energies of tritium,
helium-3, helium-4, lithium-6, and carbon-12 nuclei. \ Our results were
previously summarized in a letter publication. \ This paper provides full
details of the calculations. \ We include isospin-breaking, Coulomb effects,
and interactions up to next-to-next-to-leading order in chiral effective field theory.

\end{abstract}

\maketitle
\preprint{ }

\section{Introduction}

Lattice effective field theory combines the theoretical framework of effective
field theory with numerical lattice methods. \ In contrast with most other ab
initio methods, systematic errors are all introduced at the beginning when
defining the truncated low-energy effective theory. \ The errors can be
clearly identified as either missing operators in the lattice action, finite
volume effects, or errors from finite\ Euclidean-time extrapolation. \ Future
studies can build upon existing calculations in a straightforward manner by
including the missing operators, increasing the volume, or improving the
Euclidean-time extrapolation.

Lattice effective field theory has been used to study nuclear matter
\cite{Muller:1999cp} and neutron matter
\cite{Lee:2004qd,Lee:2004si,Abe:2007fe,Borasoy:2007vk,Epelbaum:2008vj,Wlazlowski:2009yi}%
. \ The method has also been applied to nuclei with $A\leq4$ using effective
field theory with and without pions
\cite{Borasoy:2005yc,Borasoy:2006qn,Epelbaum:2009zs}. \ A review of lattice
effective field theory calculations can be found in Ref.~\cite{Lee:2008fa}.
\ Reviews of chiral effective field theory can be found in
Ref.~\cite{vanKolck:1999mw,Bedaque:2002mn,Epelbaum:2005pn,Epelbaum:2008ga}.

In this paper we present the first lattice results for lithium-6 and carbon-12
using chiral effective field theory. \ We also present the first lattice
calculations to include isospin-breaking and Coulomb effects. \ Our results
were previously summarized in a letter publication \cite{Epelbaum:2009pd}.
\ This paper provides full details of the calculations. \ We begin by
describing the lattice interactions in chiral effective field theory appearing
at leading order, next-to-leading order, and next-to-next-to-leading order.
\ This is followed by a discussion of isospin-breaking and Coulomb
interactions. \ After this all unknown operator coefficients are fit using
low-energy scattering data. \ We then compute the energy splitting between the
triton and helium-3. \ We discuss the auxiliary-field Monte Carlo projection
method and an approximate universality of contributions from higher-order
interactions in systems with four or more nucleons. \ This is followed by
lattice results for the ground state energy of helium-4, lithium-6, and carbon-12.

\section{Leading order}

The low-energy expansion in effective field theory counts powers of the ratio
$Q$/$\Lambda$. $\ Q$ is the momentum scale associated with the mass of the
pion or external nucleon momenta, and $\Lambda$ is the momentum scale at which
the effective theory breaks down. \ At leading order (LO) in the Weinberg
power-counting scheme \cite{Weinberg:1990rz,Weinberg:1991um}, the
nucleon-nucleon effective potential contains two independent contact
interactions and instantaneous one-pion exchange. \ As in previous lattice
studies we make use of an \textquotedblleft improved\textquotedblright%
\ leading-order\ action. \ This improved leading-order action is treated
completely non-perturbatively, while higher-order interactions are included as
a perturbative expansion in powers of $Q/\Lambda$.

In our lattice calculations we use the improved LO$_{3}$ lattice action
introduced in Ref.~\cite{Epelbaum:2008vj} with spatial lattice spacing
$a=(100$~MeV$)^{-1}=1.97$~fm and temporal lattice spacing $a_{t}%
=(150$~MeV$)^{-1}=1.32$~fm. \ We take the parameter values $g_{A}=1.29$,
$\ f_{\pi}=92.2$~MeV, $m_{\pi}=m_{\pi^{0}}=134.98$~MeV. \ For the nucleon mass
we use $m=938.92$~MeV. \ Many of the calculations presented in this paper have
never been attempted before, and our choice of spatial lattice spacing is made
to optimize the efficiency of the Monte Carlo lattice calculations. \ While
$1.97$~fm is much larger than lattice spacings used lattice QCD simulations,
we should emphasize that we are not probing the quark and gluon substructure
of nucleons but rather the distribution of nucleons within nuclei. \ Our
lattice spacing corresponds with a maximum filling density of more than three
times normal nuclear matter density. \ In future studies the same systems will
also be analyzed using smaller lattice spacings.

Throughout this discussion we first present the interactions in continuum
notation and then later give the corresponding lattice operator. \ For the
continuum notation we give matrix elements for incoming and outgoing
two-nucleon momentum states. \ In the following $\vec{q}$ denotes the
$t$-channel momentum transfer. \ We use $\boldsymbol{\tau}$ to represent Pauli
matrices in isospin space and $\vec{\sigma}$ for Pauli matrices in spin space.
\ The interactions correspond with the amplitude,%
\begin{align}
\mathcal{A}\left(  V_{\text{LO}}\right)   &  =C_{S=0,I=1}f(\vec{q})\left(
\frac{1}{4}-\frac{1}{4}\vec{\sigma}_{A}\cdot\vec{\sigma}_{B}\right)  \left(
\frac{3}{4}+\frac{1}{4}\boldsymbol{\tau}_{A}\cdot\boldsymbol{\tau}_{B}\right)
\nonumber\\
&  +C_{S=1,I=0}f(\vec{q})\left(  \frac{3}{4}+\frac{1}{4}\vec{\sigma}_{A}%
\cdot\vec{\sigma}_{B}\right)  \left(  \frac{1}{4}-\frac{1}{4}\boldsymbol{\tau
}_{A}\cdot\boldsymbol{\tau}_{B}\right) \nonumber\\
&  -\left(  \frac{g_{A}}{2f_{\pi}}\right)  ^{2}\frac{\left(  \boldsymbol{\tau
}_{A}\cdot\boldsymbol{\tau}_{B}\right)  \left(  \vec{q}\cdot\vec{\sigma}%
_{A}\right)  \left(  \vec{q}\cdot\vec{\sigma}_{B}\right)  }{q^{2}+m_{\pi}^{2}%
}.
\end{align}

We use a Euclidean-time transfer-matrix lattice formalism. \ The transfer
matrix is the normal-ordered exponential of the lattice Hamiltonian,
$\colon\exp(-H\Delta t)\colon$, where $\Delta t$ equals one temporal lattice
spacing. \ We use the lattice notation adopted in several previous
publications and which is summarized in the appendix. \ Let $V_{S=0,I=1}$ be
the lattice density-density correlation for the spin-singlet isospin-triplet
channel in momentum space,%
\begin{align}
V_{S=0,I=1}(\vec{q})  &  =\frac{3}{32}:\rho^{a^{\dag},a}(\vec{q})\rho
^{a^{\dag},a}(-\vec{q}):-\frac{3}{32}:\sum_{S}\rho_{S}^{a^{\dag},a}(\vec
{q})\rho_{S}^{a^{\dag},a}(-\vec{q}):\nonumber\\
&  +\frac{1}{32}:\sum_{I}\rho_{I}^{a^{\dag},a}(\vec{q})\rho_{I}^{a^{\dag}%
,a}(-\vec{q}):-\frac{1}{32}:\sum_{S,I}\rho_{S,I}^{a^{\dag},a}(\vec{q}%
)\rho_{S,I}^{a^{\dag},a}(-\vec{q}):.
\end{align}
Let $V_{S=1,I=0}$ be the density-density correlation for the spin-triplet
isospin-singlet channel,%
\begin{align}
V_{S=1,I=0}(\vec{q})  &  =\frac{3}{32}:\rho^{a^{\dag},a}(\vec{q})\rho
^{a^{\dag},a}(-\vec{q}):+\frac{1}{32}:\sum_{S}\rho_{S}^{a^{\dag},a}(\vec
{q})\rho_{S}^{a^{\dag},a}(-\vec{q}):\nonumber\\
&  -\frac{3}{32}:\sum_{I}\rho_{I}^{a^{\dag},a}(\vec{q})\rho_{I}^{a^{\dag}%
,a}(-\vec{q}):-\frac{1}{32}:\sum_{S,I}\rho_{S,I}^{a^{\dag},a}(\vec{q}%
)\rho_{S,I}^{a^{\dag},a}(-\vec{q}):.
\end{align}
We use these functions to write the leading-order transfer matrix,%
\begin{align}
M_{\text{LO}}  &  =\colon\exp\left\{  -H_{\text{free}}\alpha_{t}-\frac
{\alpha_{t}}{L^{3}}\sum_{\vec{q}}f(\vec{q})\left[  C_{S=0,I=1}V_{S=0,I=1}%
(\vec{q})+C_{S=1,I=0}V_{S=1,I=0}(\vec{q})\right]  \right. \nonumber\\
&  \qquad+\left.  \frac{g_{A}^{2}\alpha_{t}^{2}}{8f_{\pi}^{2}q_{\pi}}%
\sum_{\substack{S_{1},S_{2},I}}\sum_{\vec{n}_{1},\vec{n}_{2}}G_{S_{1}S_{2}%
}(\vec{n}_{1}-\vec{n}_{2})\rho_{S_{1},I}^{a^{\dag},a}(\vec{n}_{1})\rho
_{S_{2},I}^{a^{\dag},a}(\vec{n}_{2})\right\}  \colon.
\end{align}
The momentum-dependent coefficient function $f(\vec{q})$ is given by
\begin{equation}
f(\vec{q})=f_{0}^{-1}\exp\left[  -b%
{\displaystyle\sum\limits_{l}}
\left(  1-\cos q_{l}\right)  \right]  ,
\end{equation}
where%
\begin{equation}
f_{0}=\frac{1}{L^{3}}\sum_{\vec{q}}\exp\left[  -b%
{\displaystyle\sum\limits_{l}}
\left(  1-\cos q_{l}\right)  \right]  .
\end{equation}
We use the value $b=0.6$, which gives approximately the correct effective
range for the two $S$-wave channels when $C_{S=0,I=1}$ and $C_{S=1,I=0}$ are
tuned to the physical $S$-wave scattering lengths.

\section{Next-to-leading order}

At next-to-leading order (NLO) the two-nucleon effective potential includes
seven contact interactions carrying two powers of momentum, corrections to the
two LO\ contact interactions, and the leading contribution from the
instantaneous two-pion exchange potential (TPEP)
\cite{Ordonez:1992xp,Ordonez:1993tn,Ordonez:1996rz,Epelbaum:1998ka,Epelbaum:1999dj}%
,%
\begin{equation}
V_{\text{NLO}}=V_{\text{LO}}+\Delta V^{(0)}+V^{(2)}+V_{\text{NLO}%
}^{\text{TPEP}}. \label{VNLO}%
\end{equation}
The tree-level amplitudes for the contact interactions are%
\begin{equation}
\mathcal{A}\left(  \Delta V^{(0)}\right)  =\Delta C+\Delta C_{I^{2}%
}\boldsymbol{\tau}_{A}\cdot\boldsymbol{\tau}_{B} \label{dV0}%
\end{equation}
and%
\begin{align}
\mathcal{A}\left(  V^{(2)}\right)   &  =C_{1}q^{2}+C_{2}k^{2}+\left(
C_{3}q^{2}+C_{4}k^{2}\right)  \left(  \vec{\sigma}_{A}\cdot\vec{\sigma}%
_{B}\right) \nonumber\\
&  +iC_{5}\frac{1}{2}\left(  \vec{\sigma}_{A}+\vec{\sigma}_{B}\right)
\cdot\left(  \vec{q}\times\vec{k}\right) \nonumber\\
&  +C_{6}\left(  \vec{q}\cdot\vec{\sigma}_{A}\right)  \left(  \vec{q}\cdot
\vec{\sigma}_{B}\right)  +C_{7}\left(  \vec{\sigma}_{A}\cdot\vec{k}\right)
\left(  \vec{\sigma}_{B}\cdot\vec{k}\right)  . \label{V2}%
\end{align}
The amplitude for the NLO two-pion exchange potential is
\cite{Friar:1994,Kaiser:1997mw}%

\begin{align}
\mathcal{A}\left[  V_{\text{NLO}}^{\text{TPEP}}\right]   &  =-\frac
{\boldsymbol{\tau}_{A}\cdot\boldsymbol{\tau}_{B}}{384\pi^{2}f_{\pi}^{4}%
}L(q)\left[  4m_{\pi}^{2}\left(  5g_{A}^{4}-4g_{A}^{2}-1\right)  +q^{2}\left(
23g_{A}^{4}-10g_{A}^{2}-1\right)  +\frac{48g_{A}^{4}m_{\pi}^{4}}{4m_{\pi}%
^{2}+q^{2}}\right] \nonumber\\
&  -\frac{3g_{A}^{4}}{64\pi^{2}f_{\pi}^{4}}L(q)\left[  \left(  \vec{q}%
\cdot\vec{\sigma}_{A}\right)  \left(  \vec{q}\cdot\vec{\sigma}_{B}\right)
-q^{2}\left(  \vec{\sigma}_{A}\cdot\vec{\sigma}_{B}\right)  \right]  ,
\label{VTPEPNLO}%
\end{align}
where%
\begin{equation}
L(q)=\frac{1}{2q}\sqrt{4m_{\pi}^{2}+q^{2}}\ln\frac{\sqrt{4m_{\pi}^{2}+q^{2}%
}+q}{\sqrt{4m_{\pi}^{2}+q^{2}}-q}. \label{Lq}%
\end{equation}

\begin{figure}[ptb]%
\centering
\includegraphics[
height=2.8245in,
width=4.0127in
]%
{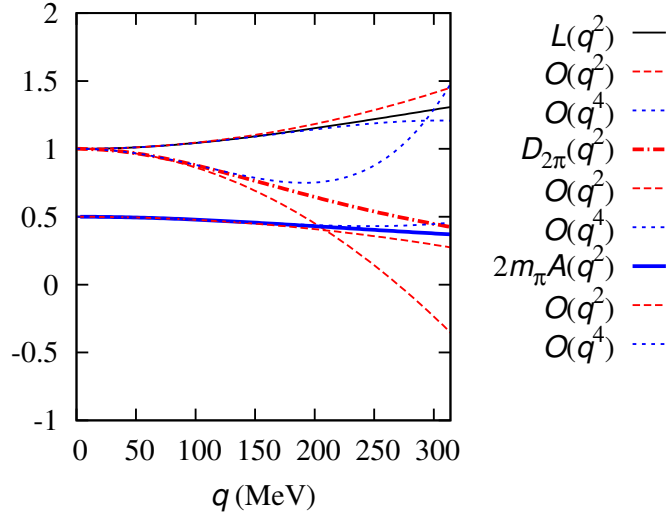}%
\caption{The functions $L(q)$, $D_{2\pi}(q)$, and $2m_{\pi}A(q)$ appearing in
the two-pion exchange potential and comparsions with their analytic expansions
up to $O(q^{2})$ and $O(q^{4})$.}%
\label{TPE}%
\end{figure}

In the lattice calculations we use a low-cutoff modification of the usual
power counting scheme. \ For nearly all $q<\Lambda$ we can expand the NLO
two-pion exchange potential in powers of $q^{2}/(4m_{\pi}^{2})$. \ This
expansion fails to converge only for values of $q$ near the cutoff scale
$\Lambda$ $\approx2.3m_{\pi}$, where the effective theory already breaks down
due to large cutoff effects. \ In Fig.~(\ref{TPE}) we show the various
functions appearing in the two-pion exchange potential and comparsions with
their analytic expansions up to $O(q^{2})$ and $O(q^{4})$. \ We show the
function $L(q)$, the dimensionless $2m_{\pi}$ pole function,%
\begin{equation}
D_{2\pi}(q^{2})=\frac{4m_{\pi}^{2}}{4m_{\pi}^{2}+q^{2}},
\end{equation}
as well as the dimensionless function $2m_{\pi}A(q)$. $\ $The function $A(q)$
appears later in our discussion, Eq.~(\ref{Aq}), in connection with the NNLO
two-pion exchange potential.%
In each case the analytic expansion approximates the full function quite well
for $q$ less than $200$~MeV. \ For our chosen lattice spacing, this covers the
entire range of validity expected for the low-energy effective theory.

Instead of retaining the full non-local structure of $V_{\text{NLO}%
}^{\text{TPEP}}$ at this lattice spacing, we simply use%
\begin{equation}
V_{\text{LO}}=V^{(0)}+V^{\text{OPEP}},
\end{equation}%
\begin{equation}
V_{\text{NLO}}=V_{\text{LO}}+\Delta V^{(0)}+V^{(2)}.
\end{equation}
Terms with up to two powers of $q$ from the momentum expansion of
$V_{\text{NLO}}^{\text{TPEP}}$ are absorbed as a redefinition of the
coefficients in $\Delta V^{(0)}$ and $V^{(2)}$.

At next-to-leading order the lattice transfer matrix is%
\begin{align}
M_{\text{NLO}}  &  =M_{\text{LO}}-\left.  \alpha_{t}\colon\left[  \Delta
V+\Delta V_{I^{2}}+V_{q^{2}}+V_{I^{2},q^{2}}+V_{S^{2},q^{2}}\right.  \right.
\nonumber\\
&  \left.  \qquad\qquad\qquad\qquad+V_{S^{2},I^{2},q^{2}}+V_{(q\cdot S)^{2}%
}+V_{I^{2},(q\cdot S)^{2}}+V_{(iq\times S)\cdot k}^{I=1}\right]  M_{\text{LO}%
}\colon\text{.}%
\end{align}
The corrections to the leading-order contact interactions are%
\begin{equation}
\Delta V=\frac{1}{2}\Delta C:\sum\limits_{\vec{n}}\rho^{a^{\dagger},a}(\vec
{n})\rho^{a^{\dagger},a}(\vec{n}):,
\end{equation}%
\begin{equation}
\Delta V_{I^{2}}=\frac{1}{2}\Delta C_{I^{2}}:\sum\limits_{\vec{n},I}\rho
_{I}^{a^{\dagger},a}(\vec{n})\rho_{I}^{a^{\dagger},a}(\vec{n}):,
\end{equation}
and the seven independent contact interactions with two derivatives are%
\begin{equation}
V_{q^{2}}=-\frac{1}{2}C_{q^{2}}:\sum\limits_{\vec{n},l}\rho^{a^{\dagger}%
,a}(\vec{n})\triangledown_{l}^{2}\rho^{a^{\dagger},a}(\vec{n}):,
\end{equation}%
\begin{equation}
V_{I^{2},q^{2}}=-\frac{1}{2}C_{I^{2},q^{2}}:\sum\limits_{\vec{n},I,l}\rho
_{I}^{a^{\dagger},a}(\vec{n})\triangledown_{l}^{2}\rho_{I}^{a^{\dagger}%
,a}(\vec{n}):,
\end{equation}%
\begin{equation}
V_{S^{2},q^{2}}=-\frac{1}{2}C_{S^{2},q^{2}}:\sum\limits_{\vec{n},S,l}\rho
_{S}^{a^{\dagger},a}(\vec{n})\triangledown_{l}^{2}\rho_{S}^{a^{\dagger}%
,a}(\vec{n}):,
\end{equation}%
\begin{equation}
V_{S^{2},I^{2},q^{2}}=-\frac{1}{2}C_{S^{2},I^{2},q^{2}}:\sum\limits_{\vec
{n},S,I,l}\rho_{S,I}^{a^{\dagger},a}(\vec{n})\triangledown_{l}^{2}\rho
_{S,I}^{a^{\dagger},a}(\vec{n}):,
\end{equation}%
\begin{equation}
V_{(q\cdot S)^{2}}=\frac{1}{2}C_{(q\cdot S)^{2}}:\sum\limits_{\vec{n}}%
\sum\limits_{S}\Delta_{S}\rho_{S}^{a^{\dagger},a}(\vec{n})\sum
\limits_{S^{\prime}}\Delta_{S^{\prime}}\rho_{S^{\prime}}^{a^{\dagger},a}%
(\vec{n}):,
\end{equation}%
\begin{equation}
V_{I^{2},(q\cdot S)^{2}}=\frac{1}{2}C_{I^{2},(q\cdot S)^{2}}:\sum
\limits_{\vec{n},I}\sum\limits_{S}\Delta_{S}\rho_{S,I}^{a^{\dagger},a}(\vec
{n})\sum\limits_{S^{\prime}}\Delta_{S^{\prime}}\rho_{S^{\prime},I}%
^{a^{\dagger},a}(\vec{n}):,
\end{equation}%
\begin{align}
V_{(iq\times S)\cdot k}^{I=1}  &  =-\frac{i}{2}C_{(iq\times S)\cdot k}%
^{I=1}\left\{  \frac{3}{4}:\sum\limits_{\vec{n},l,S,l^{\prime}}\varepsilon
_{l,S,l^{\prime}}\left[  \Pi_{l}^{a^{\dagger},a}(\vec{n})\Delta_{l^{\prime}%
}\rho_{S}^{a^{\dagger},a}(\vec{n})+\Pi_{l,S}^{a^{\dagger},a}(\vec{n}%
)\Delta_{l^{\prime}}\rho^{a^{\dagger},a}(\vec{n})\right]  :\right. \nonumber\\
&  +\left.  \frac{1}{4}:\sum\limits_{\vec{n},l,S,l^{\prime},I}\varepsilon
_{l,S,l^{\prime}}\left[  \Pi_{l,I}^{a^{\dagger},a}(\vec{n})\Delta_{l^{\prime}%
}\rho_{S,I}^{a^{\dagger},a}(\vec{n})+\Pi_{l,S,I}^{a^{\dagger},a}(\vec
{n})\Delta_{l^{\prime}}\rho_{I}^{a^{\dagger},a}(\vec{n})\right]  :\right\}  .
\end{align}
The densities, current densities, and symbols $\Delta_{l}$ and $\triangledown
_{l}^{2}$, are defined in the appendix. \ The $V_{(iq\times S)\cdot k}^{I=1}$
term eliminates lattice artifacts in the spin-triplet even-parity channels.
\ This is accomplished by projecting onto the isospin-triplet channel.

\section{Next-to-next-to-leading order}

At next-to-next-to-leading order (NNLO) there are no additional two-nucleon
contact interactions. \ The two-pion exchange potential contains a subleading
contribution,%
\begin{align}
\mathcal{A}\left[  V_{\text{NNLO}}^{\text{TPEP}}\right]   &  =-\frac
{3g_{A}^{2}}{16\pi f_{\pi}^{4}}A(q)\left(  2m_{\pi}^{2}+q^{2}\right)  \left[
2m_{\pi}^{2}\left(  2c_{1}-c_{3}\right)  -c_{3}q^{2}\right] \nonumber\\
&  -\frac{g_{A}^{2}c_{4}\left(  \boldsymbol{\tau}_{A}\cdot\boldsymbol{\tau
}_{B}\right)  }{32\pi f_{\pi}^{4}}A(q)\left(  4m_{\pi}^{2}+q^{2}\right)
\left[  \left(  \vec{q}\cdot\vec{\sigma}_{A}\right)  \left(  \vec{q}\cdot
\vec{\sigma}_{B}\right)  -q^{2}\left(  \vec{\sigma}_{A}\cdot\vec{\sigma}%
_{B}\right)  \right]  ,
\end{align}
where%
\begin{equation}
A(q)=\frac{1}{2q}\arctan\frac{q}{2m_{\pi}}. \label{Aq}%
\end{equation}
However our low-cutoff expansion in powers $q^{2}/(4m_{\pi}^{2})$ reduces the
NNLO two-pion exchange potential to a sum of contact interactions with at
least four powers of $q$. \ So in this scheme there are no additional
contributions to the two-nucleon potential at NNLO. \ The only new
contributions at NNLO are due to three-nucleon interactions,%
\begin{equation}
V_{\text{NNLO}}=V_{\text{NLO}}+V_{\text{NNLO}}^{(3N)}.
\end{equation}

\begin{figure}[ptb]%
\centering
\includegraphics[
height=1.5869in,
width=2.4742in
]%
{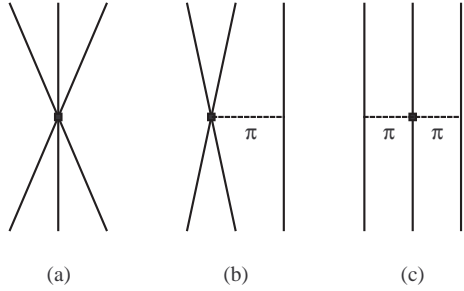}%
\caption{Three-nucleon forces at NNLO. \ Diagrams (a), (b), and (c) show the
contact potential, $V_{\text{contact}}^{(3N)}$, one-pion exchange potential
$V_{\text{OPE}}^{(3N)}$, and two-pion exchange potential $V_{\text{TPE}%
}^{(3N)}$.}%
\label{threebody}%
\end{figure}

Few-nucleon forces in chiral effective field theory beyond two nucleons were
introduced in Ref.~\cite{Weinberg:1991um}. \ In Ref.~\cite{vanKolck:1994yi} it
was shown that three-body effects first appear at next-to-next-to-leading
order (NNLO). \ The NNLO three-nucleon effective potential includes a pure
contact potential, $V_{\text{contact}}^{(3N)}$, one-pion exchange potential,
$V_{\text{OPE}}^{(3N)}$, and a two-pion exchange potential, $V_{\text{TPE}%
}^{(3N)}$,
\begin{equation}
V_{\text{NNLO}}^{(3N)}=V_{\text{contact}}^{(3N)}+V_{\text{OPE}}^{(3N)}%
+V_{\text{TPE}}^{(3N)}.
\end{equation}
The corresponding diagrams are shown in Fig.~\ref{threebody}.%

Similar to our continuum notation for two-nucleon interactions, we write the
tree-level amplitude for three-nucleon interactions with nucleons $A$, $B$,
$C$. \ We sum over all permutations $P(A,B,C)$ of the labels, and $\vec{q}%
_{A}$, $\vec{q}_{B}$, $\vec{q}_{C}$ are defined as the differences between
final and initial momenta for the respective nucleons. \ The amplitudes for
$V_{\text{contact}}^{(3N)}$ and $V_{\text{OPE}}^{(3N)}$ are
\cite{Friar:1998zt,Epelbaum:2002vt}%
\begin{equation}
\mathcal{A}\left[  V_{\text{contact}}^{(3N)}\right]  =\frac{1}{2}%
E\sum_{P(A,B,C)}\left(  \boldsymbol{\tau}_{A}\cdot\boldsymbol{\tau}%
_{B}\right)  , \label{contact_cont}%
\end{equation}%
\begin{equation}
\mathcal{A}\left[  V_{\text{OPE}}^{(3N)}\right]  =-\frac{g_{A}}{8f_{\pi}^{2}%
}D\sum_{P\left(  A,B,C\right)  }\frac{\vec{q}_{A}\cdot\vec{\sigma}_{A}}%
{q_{A}^{2}+m_{\pi}^{2}}\left(  \vec{q}_{A}\cdot\vec{\sigma}_{B}\right)
\left(  \boldsymbol{\tau}_{A}\cdot\boldsymbol{\tau}_{B}\right)  .
\label{OPE_cont}%
\end{equation}
Following the notation in Ref.~\cite{Epelbaum:2002vt}, we define dimensionless
parameters $c_{E}$ and $c_{D}$,
\begin{equation}
E=\frac{c_{E}}{f_{\pi}^{4}\Lambda_{\chi}},\quad D=\frac{c_{D}}{f_{\pi}%
^{2}\Lambda_{\chi}}\text{,}%
\end{equation}
and take $\Lambda_{\chi}=700$~MeV.

For convenience we separately label three parts of the two-pion exchange
potential$,$%
\begin{equation}
V_{\text{TPE}}^{(3N)}=V_{\text{TPE1}}^{(3N)}+V_{\text{TPE2}}^{(3N)}%
+V_{\text{TPE3}}^{(3N)}.
\end{equation}
The corresponding amplitudes are%
\begin{equation}
\mathcal{A}\left[  V_{\text{TPE1}}^{(3N)}\right]  =\frac{c_{3}}{f_{\pi}^{2}%
}\left(  \frac{g_{A}}{2f_{\pi}}\right)  ^{2}\sum_{P\left(  A,B,C\right)
}\frac{\left(  \vec{q}_{A}\cdot\vec{\sigma}_{A}\right)  \left(  \vec{q}%
_{B}\cdot\vec{\sigma}_{B}\right)  }{\left(  q_{A}^{2}+m_{\pi}^{2}\right)
\left(  q_{B}^{2}+m_{\pi}^{2}\right)  }\left(  \vec{q}_{A}\cdot\vec{q}%
_{B}\right)  \left(  \boldsymbol{\tau}_{A}\cdot\boldsymbol{\tau}_{B}\right)  ,
\label{TPE1_cont}%
\end{equation}%
\begin{equation}
\mathcal{A}\left[  V_{\text{TPE2}}^{(3N)}\right]  =-\frac{2c_{1}m_{\pi}^{2}%
}{f_{\pi}^{2}}\left(  \frac{g_{A}}{2f_{\pi}}\right)  ^{2}\sum_{P\left(
A,B,C\right)  }\frac{\left(  \vec{q}_{A}\cdot\vec{\sigma}_{A}\right)  \left(
\vec{q}_{B}\cdot\vec{\sigma}_{B}\right)  }{\left(  q_{A}^{2}+m_{\pi}%
^{2}\right)  \left(  q_{B}^{2}+m_{\pi}^{2}\right)  }\left(  \boldsymbol{\tau
}_{A}\cdot\boldsymbol{\tau}_{B}\right)  , \label{TPE2_cont}%
\end{equation}%
\begin{align}
\mathcal{A}\left[  V_{\text{TPE3}}^{(3N)}\right]   &  =\frac{c_{4}}{2f_{\pi
}^{2}}\left(  \frac{g_{A}}{2f_{\pi}}\right)  ^{2}\nonumber\\
&  \times\sum_{P\left(  A,B,C\right)  }\frac{\left(  \vec{q}_{A}\cdot
\vec{\sigma}_{A}\right)  \left(  \vec{q}_{B}\cdot\vec{\sigma}_{B}\right)
}{\left(  q_{A}^{2}+m_{\pi}^{2}\right)  \left(  q_{B}^{2}+m_{\pi}^{2}\right)
}\left[  \left(  \vec{q}_{A}\times\vec{q}_{B}\right)  \cdot\vec{\sigma}%
_{C}\right]  \left[  \left(  \boldsymbol{\tau}_{A}\times\boldsymbol{\tau}%
_{B}\right)  \cdot\boldsymbol{\tau}_{C}\right]  . \label{TPE3_cont}%
\end{align}
The constants $c_{1},c_{3},c_{4}$ parameterize the coupling of the nucleon to
two pions. \ These have been determined from fits to low-energy pion-nucleon
scattering data, and the values $c_{1}=-0.81$~GeV$^{-1}$, $c_{3}%
=-4.7$~GeV$^{-1}$, $c_{4}=3.4$~GeV$^{-1}$ are used here
\cite{Bernard:1995dp,Buettiker:1999ap}.

At next-to-next-to-leading order the lattice transfer matrix is%
\begin{equation}
M_{\text{NNLO}}=M_{\text{NLO}}-\left.  \alpha_{t}\colon\right.  \left[
V_{\text{contact}}^{(3N)}+V_{\text{OPE}}^{(3N)}+V_{\text{TPE1}}^{(3N)}%
+V_{\text{TPE2}}^{(3N)}+V_{\text{TPE3}}^{(3N)}\right]  M_{\text{LO}}:.
\end{equation}
From the constraints of isospin symmetry, spin symmetry, and Fermi statistics,
there is only one independent three-nucleon contact interaction
\cite{Bedaque:1999ve,Epelbaum:2002vt}. \ For our lattice action the contact
interaction $V_{\text{contact}}^{(3N)}$ is a product of total nucleon
densities,%
\begin{equation}
V_{\text{contact}}^{(3N)}=\frac{1}{6}D_{\text{contact}}^{(3N)}:\sum_{\vec{n}%
}\left[  \rho^{a^{\dagger},a}(\vec{n})\right]  ^{3}:\text{.}%
\end{equation}
The one-pion exchange potential $V_{\text{OPE}}^{(3N)}$ can be written as%
\begin{equation}
V_{\text{OPE}}^{(3N)}=-D_{\text{OPE}}^{(3N)}\frac{g_{A}\alpha_{t}}{2f_{\pi
}q_{\pi}}\sum_{\vec{n},S,I}\sum_{\vec{n}^{\prime},S^{\prime}}\left\langle
\Delta_{S^{\prime}}\pi_{I}^{\prime}(\vec{n}^{\prime},n_{t})\Delta_{S}\pi
_{I}^{\prime}(\vec{n},n_{t})\right\rangle :\rho_{S^{\prime},I}^{a^{\dag}%
,a}(\vec{n}^{\prime})\rho_{S,I}^{a^{\dag},a}(\vec{n})\rho^{a^{\dag},a}(\vec
{n}):\text{.}%
\end{equation}
The three two-pion exchange terms $V_{\text{TPE1}}^{(3N)},$ $V_{\text{TPE2}%
}^{(3N)},$ $V_{\text{TPE3}}^{(3N)}$ are%
\begin{align}
V_{\text{TPE1}}^{(3N)}  &  =D_{\text{TPE1}}^{(3N)}\frac{g_{A}^{2}\alpha
_{t}^{2}}{4f_{\pi}^{2}q_{\pi}^{2}}\sum_{\vec{n},S,I}\sum_{\vec{n}^{\prime
},S^{\prime}}\sum_{\vec{n}^{\prime\prime},S^{\prime\prime}}\left[
\begin{array}
[c]{c}%
\!\\
\!
\end{array}
\left\langle \Delta_{S^{\prime}}\pi_{I}^{\prime}(\vec{n}^{\prime},n_{t}%
)\Delta_{S}\pi_{I}^{\prime}(\vec{n},n_{t})\right\rangle \right. \nonumber\\
&  \times\left.  \left\langle \Delta_{S^{\prime\prime}}\pi_{I}^{\prime}%
(\vec{n}^{\prime\prime},n_{t})\Delta_{S}\pi_{I}^{\prime}(\vec{n}%
,n_{t})\right\rangle :\rho_{S^{\prime},I}^{a^{\dag},a}(\vec{n}^{\prime}%
)\rho_{S^{\prime\prime},I}^{a^{\dag},a}(\vec{n}^{\prime\prime})\rho^{a^{\dag
},a}(\vec{n}):%
\begin{array}
[c]{c}%
\!\\
\!
\end{array}
\right]  \text{,}%
\end{align}

\begin{align}
V_{\text{TPE2}}^{(3N)}  &  =D_{\text{TPE2}}^{(3N)}m_{\pi}^{2}\frac{g_{A}%
^{2}\alpha_{t}^{2}}{4f_{\pi}^{2}q_{\pi}^{2}}\sum_{\vec{n},I}\sum_{\vec
{n}^{\prime},S^{\prime}}\sum_{\vec{n}^{\prime\prime},S^{\prime\prime}}\left[
\begin{array}
[c]{c}%
\!\\
\!
\end{array}
\left\langle \Delta_{S^{\prime}}\pi_{I}^{\prime}(\vec{n}^{\prime}%
,n_{t})\square\pi_{I}^{\prime}(\vec{n},n_{t})\right\rangle \right. \nonumber\\
&  \times\left.  \left\langle \Delta_{S^{\prime\prime}}\pi_{I}^{\prime}%
(\vec{n}^{\prime\prime},n_{t})\square\pi_{I}^{\prime}(\vec{n},n_{t}%
)\right\rangle :\rho_{S^{\prime},I}^{a^{\dag},a}(\vec{n}^{\prime}%
)\rho_{S^{\prime\prime},I}^{a^{\dag},a}(\vec{n}^{\prime\prime})\rho^{a^{\dag
},a}(\vec{n}):%
\begin{array}
[c]{c}%
\!\\
\!
\end{array}
\right]  ,
\end{align}%
\begin{align}
V_{\text{TPE3}}^{(3N)}  &  =D_{\text{TPE3}}^{(3N)}\frac{g_{A}^{2}\alpha
_{t}^{2}}{4f_{\pi}^{2}q_{\pi}^{2}}\sum_{\vec{n},S_{1},S_{2},S_{3}}\sum
_{I_{1},I_{2},I_{3}}\sum_{\vec{n}^{\prime},S^{\prime}}\sum_{\vec{n}%
^{\prime\prime},S^{\prime\prime}}\left[
\begin{array}
[c]{c}%
\!\\
\!
\end{array}
\right. \nonumber\\
&  \times\left\langle \Delta_{S^{\prime}}\pi_{I_{1}}^{\prime}(\vec{n}^{\prime
},n_{t})\Delta_{S_{1}}\pi_{I_{1}}^{\prime}(\vec{n},n_{t})\right\rangle
\left\langle \Delta_{S^{\prime\prime}}\pi_{I_{2}}^{\prime}(\vec{n}%
^{\prime\prime},n_{t})\Delta_{S_{2}}\pi_{I_{2}}^{\prime}(\vec{n}%
,n_{t})\right\rangle \nonumber\\
&  \times\left.  \varepsilon_{S_{1},S_{2},S_{3}}\varepsilon_{I_{1},I_{2}%
,I_{3}}:\rho_{S^{\prime},I_{1}}^{a^{\dag},a}(\vec{n}^{\prime})\rho
_{S^{\prime\prime},I_{2}}^{a^{\dag},a}(\vec{n}^{\prime\prime})\rho
_{S_{3},I_{3}}^{a^{\dag},a}(\vec{n}):%
\begin{array}
[c]{c}%
\!\\
\!
\end{array}
\right]  \text{.}%
\end{align}
The relations between these lattice operator coefficients and the coefficients
in Eq.~(\ref{contact_cont}-\ref{TPE3_cont}) are%
\begin{equation}
D_{\text{contact}}^{(3N)}=-3E=-\frac{3c_{E}}{f_{\pi}^{4}\Lambda_{\chi}},\qquad
D_{\text{OPE}}^{(3N)}=\frac{D}{4f_{\pi}}=\frac{c_{D}}{4f_{\pi}^{3}%
\Lambda_{\chi}},
\end{equation}%
\begin{equation}
D_{\text{TPE1}}^{(3N)}=\frac{c_{3}}{f_{\pi}^{2}},\qquad D_{\text{TPE2}}%
^{(3N)}=-\frac{2c_{1}}{f_{\pi}^{2}},\qquad D_{\text{TPE3}}^{(3N)}=\frac{c_{4}%
}{2f_{\pi}^{2}}.
\end{equation}

\section{Isospin breaking and the Coulomb interaction}

In this study we include isospin-breaking terms and the Coulomb interaction.
\ Isospin breaking (IB) in effective field theory has been addressed in the
literature
\cite{vanKolck:1996rm,vanKolck:1997fu,Epelbaum:1999zn,Friar:1999zr,Walzl:2000cx,Friar:2003yv,Epelbaum:2004xf,Epelbaum:2005fd}%
. \ In the counting scheme proposed in Ref.~\cite{Epelbaum:2005fd}, the
isospin-breaking one-pion exchange interaction and Coulomb potential are
considered to be the same size as $O(Q^{2})$ corrections at NLO. \ For the
isospin-symmetric interactions we used the neutral pion mass, $m_{\pi}%
=m_{\pi^{0}}$. \ Therefore the isospin-violating one-pion exchange interaction
due to pion mass differences is%
\begin{align}
\mathcal{A}  &  \left[  V^{\text{OPEP, IB}}\right]  =-\left(  \frac{g_{A}%
}{2f_{\pi}}\right)  ^{2}\left[  \left(  \tau_{1}\right)  _{A}\left(  \tau
_{1}\right)  _{B}+\left(  \tau_{2}\right)  _{A}\left(  \tau_{2}\right)
_{B}\right] \nonumber\\
&  \times\left(  \vec{q}\cdot\vec{\sigma}_{A}\right)  \left(  \vec{q}\cdot
\vec{\sigma}_{B}\right)  \left[  \frac{1}{q^{\,2}+m_{\pi^{\pm}}^{2}}-\frac
{1}{q^{\,2}+m_{\pi^{0}}^{2}}\right]  .
\end{align}

We treat the Coulomb potential in position space with the usual $\alpha
_{\text{EM}}/r$ repulsion between protons,%
\begin{equation}
\mathcal{A}\left[  V^{\text{EM}}\right]  =\frac{\alpha_{\text{EM}}}{r}\left(
\frac{1+\tau_{3}}{2}\right)  _{A}\left(  \frac{1+\tau_{3}}{2}\right)  _{B}.
\end{equation}
However on the lattice this definition is singular for two protons on the same
lattice site. The resolution of this problem is to include a counterterm in
the form of a proton-proton contact interaction. \ For consistency we will
include all possible two-nucleon contact interactions, namely,
neutron-neutron, proton-proton, spin-singlet neutron-proton, and spin-triplet
neutron-proton. \ Since we will fit our isospin-symmetric interaction
coefficients according to neutron-proton scattering data, the two
neutron-proton contact interactions are just linear combinations of the NLO
interactions, $\Delta V$ and $\Delta V_{I^{2}}$. \ This leaves two
isospin-breaking contact interactions. \ In momentum space the amplitude for
these contact interactions are%
\begin{equation}
\mathcal{A}\left(  V_{\text{nn}}\right)  =C_{\text{nn}}\left(  \frac
{1-\tau_{3}}{2}\right)  _{A}\left(  \frac{1-\tau_{3}}{2}\right)  _{B},
\end{equation}%
\begin{equation}
\mathcal{A}\left(  V_{\text{pp}}\right)  =C_{\text{pp}}\left(  \frac
{1+\tau_{3}}{2}\right)  _{A}\left(  \frac{1+\tau_{3}}{2}\right)  _{B}.
\end{equation}

On the lattice we add these isospin-breaking terms to the NLO\ transfer
matrix,%
\begin{equation}
M_{\text{NLO}}\rightarrow M_{\text{NLO,IB}},
\end{equation}
where%
\begin{equation}
M_{\text{NLO,IB}}=M_{\text{NLO}}-\left.  \alpha_{t}\colon\left[
V^{\text{OPEP, IB}}+V_{\text{nn}}+V_{\text{pp}}\right]  \right.  M_{\text{LO}%
}\colon.
\end{equation}
The isospin-breaking one-pion exchange operator is%
\begin{align}
V^{\text{OPEP, IB}}  &  =-\frac{g_{A}^{2}\alpha_{t}}{8f_{\pi}^{2}}\nonumber\\
&  \times\sum_{\substack{I=1,2}}\sum_{\substack{S_{1},S_{2}}}\sum_{\vec{n}%
_{1},\vec{n}_{2}}\rho_{S_{1},I}^{a^{\dag},a}(\vec{n}_{1})\rho_{S_{2}%
,I}^{a^{\dag},a}(\vec{n}_{2})\left[  \frac{G_{S_{1}S_{2}}(\vec{n}_{1}-\vec
{n}_{2},m_{\pi^{\pm}})}{q_{\pi}(m_{\pi^{\pm}})}-\frac{G_{S_{1}S_{2}}(\vec
{n}_{1}-\vec{n}_{2},m_{\pi^{0}})}{q_{\pi}(m_{\pi^{0}})}\right]  .
\end{align}
The Coulomb interaction operator is%
\begin{equation}
V^{\text{EM}}=\frac{1}{2}\alpha_{\text{EM}}\colon\sum_{\vec{n}_{1},\vec{n}%
_{2}}\frac{1}{r(\vec{n}_{1}-\vec{n}_{2})}\left[  \frac{1}{2}\rho^{a^{\dag}%
,a}(\vec{n}_{1})+\frac{1}{2}\rho_{I=3}^{a^{\dag},a}(\vec{n}_{1})\right]
\left[  \frac{1}{2}\rho^{a^{\dag},a}(\vec{n}_{2})+\frac{1}{2}\rho
_{I=3}^{a^{\dag},a}(\vec{n}_{2})\right]  \colon,
\end{equation}
where $r$ is the distance on the lattice. \ We take the value of $r$ at the
origin to be $1/2$,%
\begin{equation}
r(\vec{n})=\max\left(  \frac{1}{2},\left\vert \vec{n}\right\vert \right)  .
\end{equation}
This convention choice has no observable effect since we also have a
proton-proton contact interaction which is fitted to proton-proton scattering
data. \ The proton-proton contact operator is
\begin{equation}
V_{\text{pp}}=\frac{1}{2}C_{\text{pp}}\colon\sum_{\vec{n}}\left[  \frac{1}%
{2}\rho^{a^{\dag},a}(\vec{n})+\frac{1}{2}\rho_{I=3}^{a^{\dag},a}(\vec
{n})\right]  \left[  \frac{1}{2}\rho^{a^{\dag},a}(\vec{n})+\frac{1}{2}%
\rho_{I=3}^{a^{\dag},a}(\vec{n})\right]  \colon,
\end{equation}
and the neutron-neutron contact operator is%
\begin{equation}
V_{\text{nn}}=\frac{1}{2}C_{\text{nn}}\colon\sum_{\vec{n}}\left[  \frac{1}%
{2}\rho^{a^{\dag},a}(\vec{n})-\frac{1}{2}\rho_{I=3}^{a^{\dag},a}(\vec
{n})\right]  \left[  \frac{1}{2}\rho^{a^{\dag},a}(\vec{n})-\frac{1}{2}%
\rho_{I=3}^{a^{\dag},a}(\vec{n})\right]  \colon.
\end{equation}

\section{Lattice artifacts}

In this section we discuss the relative size of lattice artifacts produced by
lattice regularization. \ We start with lattice artifacts that break
rotational invariance. \ Lattice regularization reduces the full
three-dimensional rotational group down to the cubic subroup. \ Lattice
operators that break rotational invariance first appear at $O(Q^{2})$. \ These
include local two-nucleon operators with amplitude proportional to%
\begin{equation}
\sum_{l=1,2,3}q_{l}^{2}\left(  \sigma_{A}\right)  _{l}\left(  \sigma
_{B}\right)  _{l}. \label{local_rot_breaking}%
\end{equation}
and%
\begin{equation}
\left(  \boldsymbol{\tau}_{A}\cdot\boldsymbol{\tau}_{B}\right)  \sum
_{l=1,2,3}q_{l}^{2}\left(  \sigma_{A}\right)  _{l}\left(  \sigma_{B}\right)
_{l}.
\end{equation}
These operators contain terms with total spin equal to zero, two, and four.
\ The spin-zero part of these operators do not break rotational invariance and
are already included in our set of $O(Q^{2})$ local operators at NLO. \ The
spin-two and spin-four parts of these operators make contributions to spin-two
and spin-four transition matrix elements. \ For example they generate an
unphysical mixing between the $^{3}S_{1}$-$^{3}D_{1}$ channel and the
$^{3}D_{3}$-$^{3}G_{3}$ channel. \ In all applications discussed here,
however, we compute matrix elements of operators sandwiched between states
with definite and equal values for total spin. \ Hence the contribution of the
spin-two and spin-four operators must be quadratic or higher. \ The net result
is that these effects appear at $O(Q^{4})$. \ They should be included in
analyses which consider corrections up to N$^{3}$LO.

In addition to local terms, there are also non-local lattice artifacts
associated with the one-pion exchange potential. \ These include $O(Q^{2})$
terms from the gradient coupling of the pion,%
\begin{equation}
\frac{\boldsymbol{\tau}_{A}\cdot\boldsymbol{\tau}_{B}}{q^{2}+m_{\pi}^{2}%
}\left[  \left(  \vec{q}\cdot\vec{\sigma}_{A}\right)  \sum_{l=1,2,3}q_{l}%
^{3}\left(  \sigma_{B}\right)  _{l}+\left(  \vec{q}\cdot\vec{\sigma}%
_{B}\right)  \sum_{l=1,2,3}q_{l}^{3}\left(  \sigma_{A}\right)  _{l}\right]  ,
\label{OPE_artifact1}%
\end{equation}
and the pion propagator,%
\begin{equation}
\left(  \boldsymbol{\tau}_{A}\cdot\boldsymbol{\tau}_{B}\right)  \left(
\vec{q}\cdot\vec{\sigma}_{A}\right)  \left(  \vec{q}\cdot\vec{\sigma}%
_{B}\right)  \frac{\sum_{l=1,2,3}q_{l}^{4}}{\left(  q^{2}+m_{\pi}^{2}\right)
^{2}}. \label{OPE_artifact2}%
\end{equation}
Once again the spin-two and spin-four parts of these operators appear only
quadratically when computing matrix elements of operators sandwiched between
states with definite and equal values of total spin.

The spin-zero parts of the non-local operators in Eq.~(\ref{OPE_artifact1}%
-\ref{OPE_artifact2}) are lattice artifacts which break chiral symmetry.
\ When $q<m_{\pi}$ these operators are similar to the local $O(Q^{2})$ terms
we discussed at NLO. \ However for $m_{\pi}<q<\Lambda$ the non-locality of
these lattice artifacts becomes apparent. \ As we will see later in our
discussion of $^{3}S_{1}$-$^{3}D_{1}$ mixing, there seems to be some signal of
these artifacts in the mixing angle. \ The non-local $O(Q^{2})$ effects can be
removed in future lattice studies using an $O(a^{2})$-improved pion lattice
propagator and $O(a^{2})$-improved gradient coupling of the pion to the
nucleon. \ Similar non-local corrections to the one-pion exchange potential
are generated at $O(\alpha_{t}Q^{2}/m)$ by the nonzero temporal lattice
spacing. \ In this case, however, the effects are numerically negligible due
to our small value for the temporal lattice spacing, $a_{t}=(150$ MeV$)^{-1}$.
\ This has been checked explicitly by comparing nucleon-nucleon lattice
scattering data for several different temporal lattice spacings.

\section{Results for nucleon-nucleon scattering}

We measure phase shifts and mixing angles using the spherical wall method
\cite{Borasoy:2007vy}. \ This consists of imposing a hard spherical wall
boundary on the relative separation between the two nucleons at some chosen
radius $R_{\text{wall}}$. \ Scattering phase shifts are determined from the
energies of the spherical standing waves, and mixing angles are extracted from
projections onto spherical harmonics. \ For neutron-neutron scattering and
neutron-proton scattering, the asymptotic radial dependence for momentum $p$
and orbital angular momentum $L$ is%
\begin{equation}
u_{L}^{(p)}(r)=r\cdot R_{L}^{(p)}(r)\propto\cot\delta_{L}(p)S_{L}%
(pr)+C_{L}(pr),
\end{equation}
where $R_{L}^{(p)}(r)$ is the radial wavefunction and $S_{L}$ and $C_{L}$ are
Ricatti-Bessel functions of the first and second kind. \ For proton-proton
scattering, however, the long-range electrostatic potential requires that we
use Coulomb wavefunctions. \ We replace $S_{L}(pr)$ by $F_{L}(\eta,pr)$ and
replace $C_{L}(pr)$ by $G_{L}(\eta,pr)$, where%
\begin{equation}
\eta=\frac{\alpha_{\text{EM}}m}{2p},
\end{equation}%
\begin{equation}
F_{L}(\eta,pr)=(pr)^{L+1}e^{-ipr}c_{L}(\eta)\,_{1}F_{1}(L+1-i\eta,2L+2,2ipr),
\end{equation}%
\begin{equation}
G_{L}(\eta,pr)=\frac{\left(  2i\right)  ^{2L+1}(pr)^{L+1}e^{-ipr}%
\Gamma(L+1-i\eta)}{\Gamma(2L+2)c_{L}(\eta)}U(L+1-i\eta,2L+2,2ipr)+iF_{L}%
(\eta,pr),
\end{equation}
and
\begin{equation}
c_{L}(\eta)=\frac{2^{L}e^{-\pi\eta/2}\left\vert \Gamma(L+1+i\eta)\right\vert
}{\Gamma(2L+2)}.
\end{equation}
The function $_{1}F_{1}$ is Kummer's confluent hypergeometric function of the
first kind, and the function $U$ is Kummer's confluent hypergeometric function
of the second kind.

\begin{figure}[ptb]%
\centering
\includegraphics[
height=3.4247in,
width=4.8888in
]%
{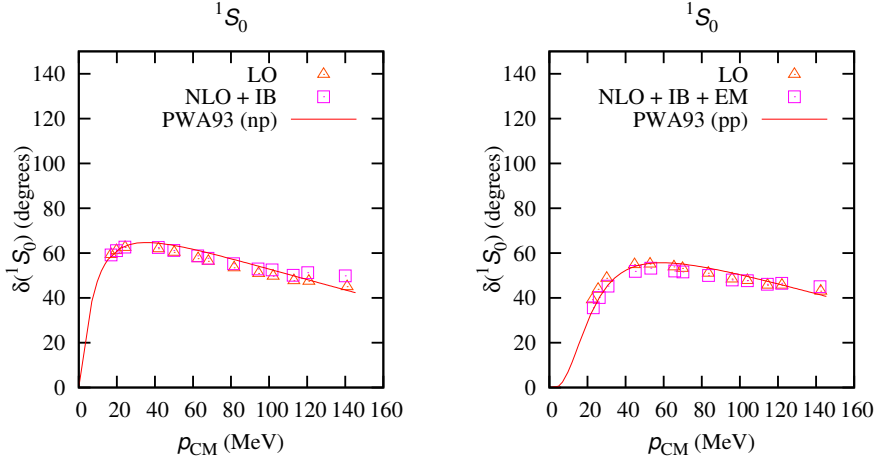}%
\caption{$^{1}S_{0}$ neutron-proton and proton-proton phase shifts versus
center of mass momentum.}%
\label{1s0_b6_100_150}%
\end{figure}

\begin{figure}[ptb]%
\centering
\includegraphics[
height=2.6801in,
width=2.4154in
]%
{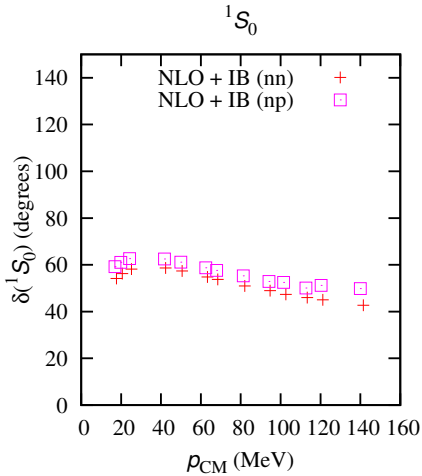}%
\caption{Comparison of the $^{1}S_{0}$ neutron-neutron and neutron-proton
phase shifts versus center of mass momentum.}%
\label{1s0_b6_100_150_nn_np}%
\end{figure}

In the following plots we show lattice scattering data for spatial lattice
spacing $a=(100$~MeV$)^{-1}$ and temporal lattice spacing $a_{t}%
=(150$~MeV$)^{-1}$. \ The $^{1}S_{0}$ neutron-proton and proton-proton phase
shifts are shown in Fig.~\ref{1s0_b6_100_150}. \ For comparison we show
partial wave results from Ref.~\cite{Stoks:1993tb}. \ We see that the
agreement is quite good for center of mass momenta up to $150$~MeV. \ To
constrain the neutron-neutron contact interaction, $C_{\text{nn}}$, we use the
neutron-neutron scattering length, which we take to be $-18$~fm with an
uncertainty of $\pm1$~fm
\cite{GonzalezTrotter:1999zt,Huhn:2000a,GonzalezTrotter:2006a,Chen:2008a}.
\ In Fig.~\ref{1s0_b6_100_150_nn_np} we show a comparison of the $^{1}S_{0}$
neutron-neutron and neutron-proton phase shifts as calculated on the lattice.%


\begin{figure}[ptb]%
\centering
\includegraphics[
height=3.4247in,
width=4.8879in
]%
{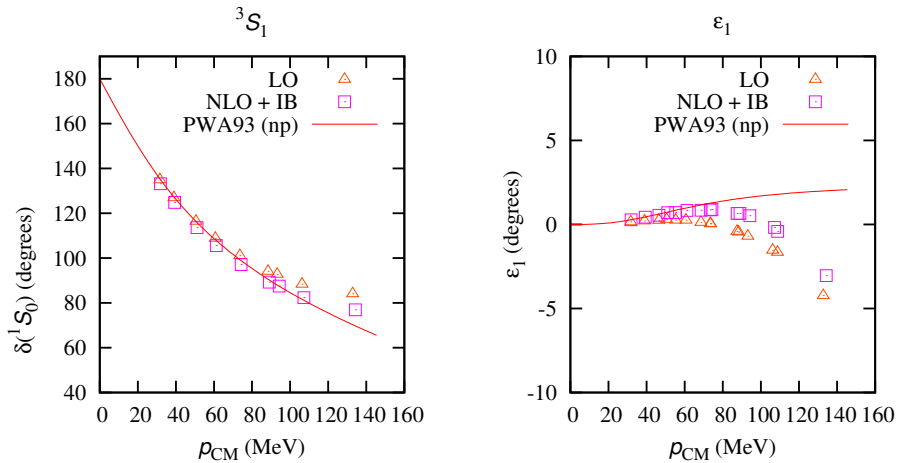}%
\caption{$^{3}S_{1}$ neutron-proton phase shift and $^{3}S_{1}$-$^{3}D_{1}$
mixing angle versus center of mass momentum.}%
\label{3s1_e1_b6_100_150}%
\end{figure}

In Fig.~\ref{3s1_e1_b6_100_150} we plot the $^{3}S_{1}$ phase shift and
$^{3}S_{1}$-$^{3}D_{1}$ mixing angle $\varepsilon_{1}$ using the Stapp
parameterization \cite{Stapp:1956mz}. \ The agreement with the results of the
Nijmegen PWA \cite{Stoks:1993tb} for the $^{3}S_{1}$ partial wave is good up
to $150$~MeV. \ The mixing angle is good at low momenta, but deviations appear
at higher momenta. \ This discrepancy is likely due to lattice artifacts such
as the terms previously discussed in Eq.~(\ref{OPE_artifact1}%
-\ref{OPE_artifact2}) as well as the contribution of higher-order
interactions. \ In future work some improvement may be possible using an
$O(a^{2})$-improved pion lattice propagator and $O(a^{2})$-improved gradient
coupling of the pion to the nucleon. \ Nonetheless the physics of $^{3}S_{1}%
$-$^{3}D_{1}$ mixing appears correct at low energies. \ This we can test by
computing the quadrupole moment of the deuteron. \ With no additional free
parameters to tune we find $0.22$~fm$^{2}$ at leading order and $0.29$%
~fm$^{2}$ at next-to-leading order with isospin-breaking contributions. \ The
quadrupole moment is related to the strength of the mixing angle at low
momenta. \ We estimate an $8\%$ uncertainty in fitting the mixing angle in
that regime, and so our result for the quadrupole moment with error bars is
$0.29(2)$~fm$^{2}$. \ This agrees well with the physical value of
$0.286$~fm$^{2}$.%

\begin{figure}[ptb]%
\centering
\includegraphics[
height=5.3748in,
width=4.8879in
]%
{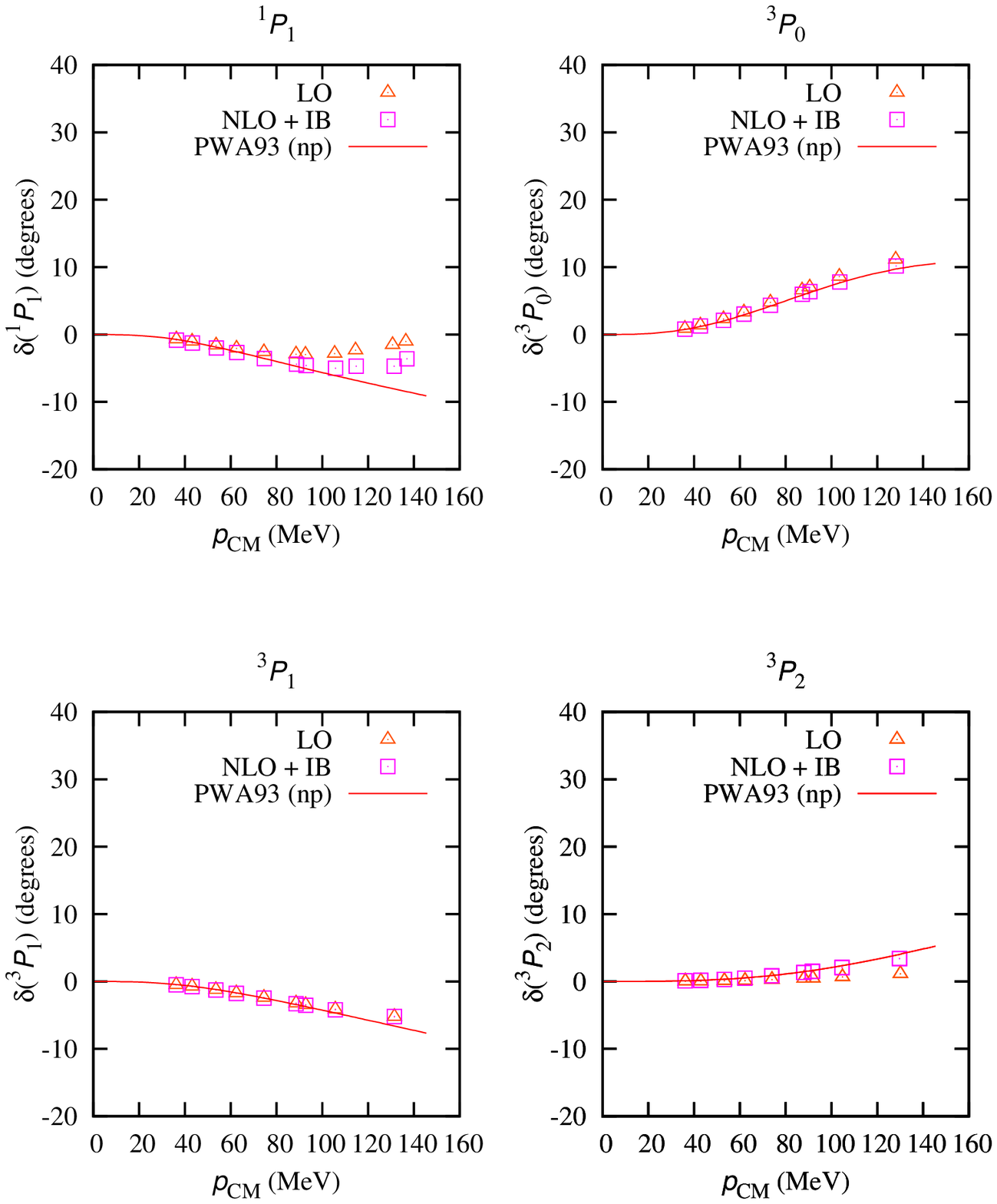}%
\caption{$^{1}P_{1}$, $^{3}P_{0}$, $^{3}P_{1}$, and $^{3}P_{2}$ neutron-proton
phase shifts versus center of mass momentum.}%
\label{pwave_b6_100_150}%
\end{figure}

In Fig.~\ref{pwave_b6_100_150} we show results for neutron-proton scattering
in the $^{1}P_{1}$, $^{3}P_{0}$, $^{3}P_{1}$, and $^{3}P_{2}$ channels. \ In
all cases the comparison with physical data \cite{Stoks:1993tb} is good up to
center of mass momenta of $150$~MeV.%

\section{Energy splitting between triton and helium-3}

The three-nucleon system is small enough that we can use iterative
sparse-matrix eigenvector methods to compute energy levels on cubic periodic
lattices. \ We fix the coefficient $c_{E}$ as a function of $c_{D}$ by
matching the physical triton energy at infinite volume, $-8.48$~MeV. \ We
consider cubes with side lengths $L$ up to $16$~fm and extract the infinite
volume limit using the asymptotic result \cite{Luscher:1985dn},%
\begin{equation}
E(L)=E(\infty)-\frac{C}{L}e^{-L/L_{0}}+O\left(  e^{-\sqrt{2}L/L_{0}}\right)
\text{.} \label{E_L}%
\end{equation}
The value of $c_{D}$ is determined from a second observable such as the
spin-doublet nucleon-deuteron scattering phase shifts. \ It turns out however
that the spin-doublet nucleon-deuteron scattering phase shift provides only a
mild constraint on $c_{D}$, namely that $c_{D}\sim O(1).$ \ Currently we are
investigating other methods for constraining $c_{D}$, including one recent
suggestion to determine $c_{D}$ from the triton beta decay rate
\cite{Gazit:2008ma}. \ In this analysis we simply use the estimate $c_{D}\sim
O(1)$ and check the dependence of observables upon $c_{D}$.

\begin{figure}[ptb]%
\centering
\includegraphics[
height=2.527in,
width=2.898in
]%
{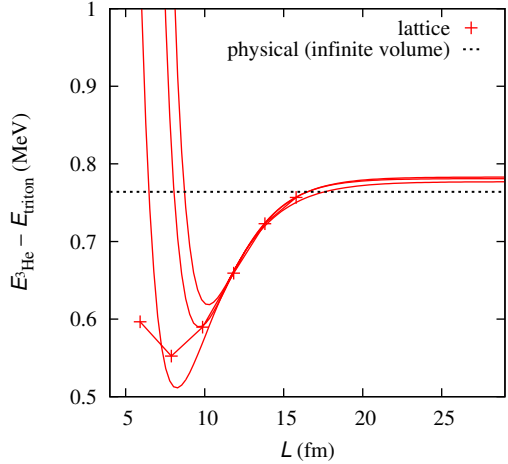}%
\caption{The energy difference between helium-3 and the triton versus periodic
cube length.}%
\label{triton_he3_difference}%
\end{figure}

Although the triton energy at infinite volume is used to set the unknown
coefficient $c_{E}$, the energy splitting between helium-3 and the triton is a
testable prediction. \ The energy difference between helium-3 and the triton
is plotted in Fig. \ref{triton_he3_difference} as a function of cube length.
\ We show several different asymptotic fits using Eq.~(\ref{E_L}) and
different subsets of data points. \ To the order at which we are working there
is no dependence of the energy splitting upon the value of $c_{D}$. \ Our
calculations at next-to-next-to-leading order give a value of $0.780$~MeV with
an infinite-volume extrapolation error of $\pm0.003$~MeV. \ To estimate other
errors we take into account an uncertainty of $\pm1$~fm in the neutron
scattering length and a $5\%$ relative uncertainty in our lattice fit of the
splitting between neutron-proton and proton-proton phase shifts at low
energies. \ Our final result for the energy splitting with error bars is then
$0.78(5)$~MeV. \ This agrees well with the experimental value of $0.76$~MeV.%

\section{Higher-order interactions}

\begin{figure}[ptb]%
\centering
\includegraphics[
height=2.2416in,
width=5.3004in
]%
{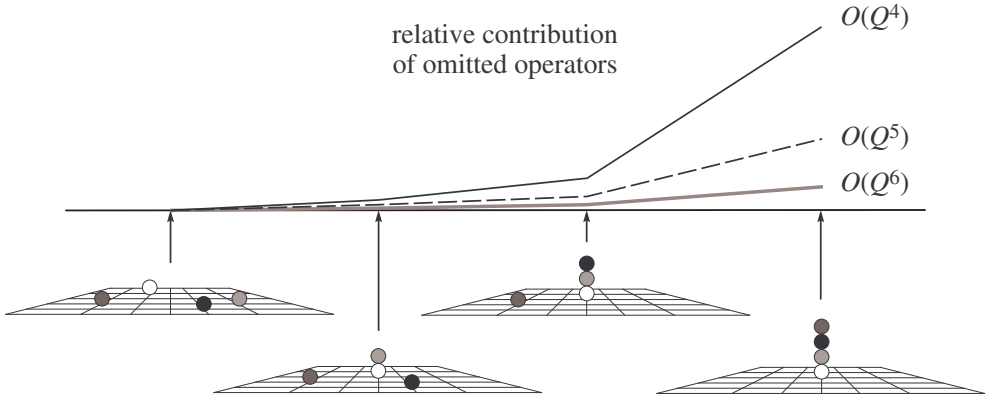}%
\caption{Sketch of the relative contribution of omitted operators at
$O(Q^{4})$, $O(Q^{5})$, $O(Q^{6})$ for various nucleon configurations. \ The
relative contribution is dominated by the last case where four nucleons are
close together.}%
\label{errors}%
\end{figure}

In this analysis we include all operators up to next-to-next-to-leading order.
\ Some residual error is expected from omitted higher-order interactions
starting at $O(Q^{4})$. \ The size of the error depends on the momentum scale
probed by the physical system of interest. \ For well-separated low-momentum
nucleons no significant deviation should occur. \ For two nucleons in close
proximity the systematic error should also remain very small. \ The properties
of the deuteron and soft nucleon-deuteron scattering are both accurately
reproduced \cite{Epelbaum:2009zs}. \ For three nucleons in close proximity the
error increases a bit more, and for a tight\ cluster of four nucleons it
increases further. \ We stop at four nucleons since a localized collection of
five or more nucleons with no relative orbital angular momentum is forbidden
by Fermi statistics. \ The expected trend for systematic errors is sketched
qualitatively in Fig.~\ref{errors}.%

As the sketch suggests, the relative contribution is likely dominated by the
last case where four nucleons are close together. \ If this hypothesis is
correct then the contribution of higher-order operators to low-energy
phenomena should be approximately universal. \ Different higher-order
operators produce roughly the same effect on low-energy data. \ This situation
is analogous to the difficulty one finds in resolving the value of $c_{D}$
from low-energy three-nucleon data. \ One useful consequence of this
universality is that most of the residual error can be cancelled by adjusting
the coefficient of an effective four-nucleon contact term,%
\begin{equation}
V_{\text{effective}}^{(4N)}=\frac{1}{24}D_{\text{effective}}^{(4N)}:\sum
_{\vec{n}}\left[  \rho^{a^{\dagger},a}(\vec{n})\right]  ^{4}:\text{.}%
\end{equation}
This effective four-nucleon contact interaction should not be confused with
the four-nucleon contact interaction that appears at $O(Q^{6})$. \ We are not
suggesting a rearrangement of power counting in chiral effective field theory.
\ We are simply taking advantage of the expected universality of missing
higher-order interactions. Later in our discussion we present results which
test and appear to confirm this universality hypothesis.

The inclusion of $V_{\text{effective}}^{(4N)}$ provides an opportunity to
resolve another related issue that was noted in earlier lattice calculations.
\ Let $\left\vert 4N_{\text{one-site}}\right\rangle $ be a configuration of
four nucleons on a single lattice site,%
\begin{equation}
\left\vert 4N_{\text{one-site}}\right\rangle =a_{0,0}^{\dagger}(\vec
{n})a_{1,0}^{\dagger}(\vec{n})a_{0,1}^{\dagger}(\vec{n})a_{1,1}^{\dagger}%
(\vec{n})\left\vert 0\right\rangle .
\end{equation}
The potential energy of this configuration is dependent upon the three-nucleon
contact operator and the local part of the three-nucleon one-pion-exchange
interaction,%
\begin{equation}
\left\langle 4N_{\text{one-site}}\right\vert V_{\text{contact}}^{(3N)}%
\left\vert 4N_{\text{one-site}}\right\rangle =4D_{\text{contact}}^{(3N)},
\label{contact_3_4}%
\end{equation}%
\begin{equation}
\left\langle 4N_{\text{one-site}}\right\vert V_{\text{OPE}}^{(3N)}\left\vert
4N_{\text{one-site}}\right\rangle =12\frac{g_{A}\alpha_{t}}{f_{\pi}q_{\pi}%
}D_{\text{OPE}}^{(3N)}\sum\limits_{S}G_{SS}(\vec{0}). \label{OPE_3_4}%
\end{equation}
If $D_{\text{contact}}^{(3N)}$ or $D_{\text{OPE}}^{(3N)}$ is sufficiently
large and negative, a clustering instability can be produced in systems with
four or more nucleons. \ This is a lattice artifact that appears on coarse
lattices \cite{Lee:2005xy}, and is similar to the clustering instability found
with point-like two-nucleon contact interactions \cite{Borasoy:2006qn}. \ That
problem was solved by using improved lattice actions with operator smearing.
\ An analogous technique could be adopted for the three-nucleon interactions.
\ In Ref.~\cite{Epelbaum:2009zs}, however, a different approach was used. \ In
that analysis the temporal lattice spacing was adjusted to ensure that the
size of the cutoff-dependent three-nucleon operator coefficients were small.

In this study we use a simpler and more direct technique. \ Let us define%
\begin{align}
D_{\text{effective}}^{\prime(4N)}  &  =\left\langle 4N_{\text{one-site}%
}\right\vert V_{\text{contact}}^{(3N)}+V_{\text{OPE}}^{(3N)}%
+V_{\text{effective}}^{(4N)}\left\vert 4N_{\text{one-site}}\right\rangle
\nonumber\\
&  =4D_{\text{contact}}^{(3N)}+12D_{\text{OPE}}^{(3N)}\frac{g_{A}\alpha_{t}%
}{f_{\pi}q_{\pi}}\sum\limits_{S}G_{SS}(\vec{0})+D_{\text{effective}}^{(4N)}.
\end{align}
The problem is that the local three-nucleon terms induce an effect much the
same as a four-nucleon contact interaction, and quite possibly a strong
four-nucleon interaction. \ To remedy this we treat $D_{\text{effective}%
}^{(4N)}$ as a bare counterterm that removes the dependence on
$D_{\text{contact}}^{(3N)}$ and $D_{\text{OPE}}^{(3N)}$. \ In the following we
express all lattice results in terms of the renormalized coupling
$D_{\text{effective}}^{\prime(4N)}.$

\section{Auxiliary fields and projection Monte Carlo}

For systems with more than three nucleons, sparse-matrix calculations using
the lattice transfer matrix are not practical at large volumes. \ Instead we
use projection Monte Carlo with auxiliary fields. \ The auxiliary-field
transfer matrix for the LO$_{3}$ action requires sixteen auxiliary fields.
\ One auxiliary field is associated with the total nucleon density
$N^{\dagger}N$, three fields for the spin density $N^{\dagger}\vec{\sigma}N$,
three fields for the isospin density $N^{\dagger}\boldsymbol{\tau}N$, and nine
fields for the spin-isospin density $N^{\dagger}\vec{\sigma}\boldsymbol{\tau
}N$. \ Let us define $M^{(n_{t})}(\pi_{I}^{\prime},s,s_{S},s_{I},s_{S,I})$ as
the leading-order\ auxiliary-field transfer matrix at time step $n_{t}$,%
\begin{align}
M^{(n_{t})}(\pi_{I}^{\prime},s,s_{S},s_{I},s_{S,I})  &  =\colon\exp\left\{
-H_{\text{free}}\alpha_{t}-\frac{g_{A}\alpha_{t}}{2f_{\pi}\sqrt{q_{\pi}}}%
{\displaystyle\sum_{\vec{n},S,I}}
\Delta_{S}\pi_{I}^{\prime}(\vec{n},n_{t})\rho_{S,I}^{a^{\dag},a}(\vec
{n})\right. \nonumber\\
&  +\frac{1}{4}\sqrt{\left(  -3C_{S=0,I=1}-3C_{S=1,I=0}\right)  \alpha_{t}%
}\sum_{\vec{n}}s(\vec{n},n_{t})\rho^{a^{\dag},a}(\vec{n})\nonumber\\
&  +\frac{i}{4}\sqrt{\left(  -3C_{S=0,I=1}+C_{S=1,I=0}\right)  \alpha_{t}}%
\sum_{\vec{n},S}s_{S}(\vec{n},n_{t})\rho_{S}^{a^{\dag},a}(\vec{n})\nonumber\\
&  +\frac{i}{4}\sqrt{\left(  C_{S=0,I=1}-3C_{S=1,I=0}\right)  \alpha_{t}}%
\sum_{\vec{n},I}s_{I}(\vec{n},n_{t})\rho_{I}^{a^{\dag},a}(\vec{n})\nonumber\\
&  \left.  +\frac{i}{4}\sqrt{\left(  -C_{S=0,I=1}-C_{S=1,I=0}\right)
\alpha_{t}}\sum_{\vec{n},S,I}s_{S,I}(\vec{n},n_{t})\rho_{S,I}^{a^{\dag}%
,a}(\vec{n})\right\}  \colon.
\end{align}
We can write $M_{\text{LO}}$ as the normalized integral%
\begin{equation}
M_{\text{LO}}=\frac{%
{\displaystyle\int}
D\pi_{I}^{\prime}DsDs_{S}Ds_{I}Ds_{S,I}\;e^{-S_{\pi\pi}^{(n_{t})}%
-S_{ss}^{(n_{t})}}M^{(n_{t})}(\pi_{I}^{\prime},s,s_{S},s_{I},s_{S,I})}{%
{\displaystyle\int}
D\pi_{I}^{\prime}DsDs_{S}Ds_{I}Ds_{S,I}\;e^{-S_{\pi\pi}^{(n_{t})}%
-S_{ss}^{(n_{t})}}}, \label{LOaux}%
\end{equation}
where $S_{\pi\pi}^{(n_{t})}$ is the piece of the instantaneous pion action at
time step $n_{t}$,%
\begin{equation}
S_{\pi\pi}^{(n_{t})}(\pi_{I}^{\prime})=\frac{1}{2}\sum_{\vec{n},I}\pi
_{I}^{\prime}(\vec{n},n_{t})\pi_{I}^{\prime}(\vec{n},n_{t})-\frac{\alpha_{t}%
}{q_{\pi}}\sum_{\vec{n},I,l}\pi_{I}^{\prime}(\vec{n},n_{t})\pi_{I}^{\prime
}(\vec{n}+\hat{l},n_{t}),
\end{equation}
and $S_{ss}^{(n_{t})}$ is the auxiliary-field action at time step $n_{t}$,%
\begin{align}
S_{ss}^{(n_{t})}  &  =\frac{1}{2}\sum_{\vec{n},\vec{n}^{\prime}}s(\vec
{n},n_{t})f^{-1}(\vec{n}-\vec{n}^{\prime})s(\vec{n}^{\prime},n_{t})+\frac
{1}{2}\sum_{\vec{n},\vec{n}^{\prime},S}s_{S}(\vec{n},n_{t})f^{-1}(\vec{n}%
-\vec{n}^{\prime})s_{S}(\vec{n}^{\prime},n_{t})\nonumber\\
&  +\frac{1}{2}\sum_{\vec{n},\vec{n}^{\prime},I}s_{I}(\vec{n},n_{t}%
)f^{-1}(\vec{n}-\vec{n}^{\prime})s_{I}(\vec{n}^{\prime},n_{t})+\frac{1}{2}%
\sum_{\vec{n},\vec{n}^{\prime},S,I}s_{S,I}(\vec{n},n_{t})f^{-1}(\vec{n}%
-\vec{n}^{\prime})s_{S,I}(\vec{n}^{\prime},n_{t}),
\end{align}
with
\begin{equation}
f^{-1}(\vec{n}-\vec{n}^{\prime})=\frac{1}{L^{3}}\sum_{\vec{q}}\frac{1}%
{f(\vec{q})}e^{-i\vec{q}\cdot(\vec{n}-\vec{n}^{\prime})}\text{.}%
\end{equation}

The contributions from NLO, NNLO, isospin-breaking, and Coulomb interactions
are treated using perturbation theory. \ This is done by including external
sources coupled to densities and current densities. \ Let us define%
\begin{equation}
M_{\text{LO}}(\varepsilon)=\frac{%
{\displaystyle\int}
D\pi_{I}^{\prime}DsDs_{S}Ds_{I}Ds_{S,I}\;e^{-S_{\pi\pi}^{(n_{t})}%
-S_{ss}^{(n_{t})}}M^{(n_{t})}(\pi_{I}^{\prime},s,s_{S},s_{I},s_{S,I}%
,\varepsilon)}{%
{\displaystyle\int}
D\pi_{I}^{\prime}DsDs_{S}Ds_{I}Ds_{S,I}\;e^{-S_{\pi\pi}^{(n_{t})}%
-S_{ss}^{(n_{t})}}},
\end{equation}
where%
\begin{align}
&  M^{(n_{t})}(\pi_{I}^{\prime},s,s_{S},s_{I},s_{S,I},\varepsilon)\nonumber\\
&  =:M^{(n_{t})}(\pi_{I}^{\prime},s,s_{S},s_{I},s_{S,I})\exp\left[
U^{(n_{t})}(\varepsilon)+U_{I^{2}}^{(n_{t})}(\varepsilon)\right]  :.
\end{align}
The isospin-independent couplings are%
\begin{align}
U^{(n_{t})}(\varepsilon)  &  =\sum_{\vec{n}}\varepsilon_{\rho}(\vec{n}%
,n_{t})\rho^{a^{\dagger},a}(\vec{n})+\sum_{\vec{n},S}\varepsilon_{\rho_{S}%
}(\vec{n},n_{t})\rho_{S}^{a^{\dagger},a}(\vec{n})+\sum_{\vec{n},S}%
\varepsilon_{\Delta_{S}\rho}(\vec{n},n_{t})\Delta_{S}\rho^{a^{\dagger},a}%
(\vec{n})\nonumber\\
&  +\sum_{\vec{n},S,S^{\prime}}\varepsilon_{\Delta_{S}\rho_{S^{\prime}}}%
(\vec{n},n_{t})\Delta_{S}\rho_{S^{\prime}}^{a^{\dagger},a}(\vec{n})+\sum
_{\vec{n},l}\varepsilon_{\triangledown_{l}^{2}\rho}(\vec{n},n_{t}%
)\triangledown_{l}^{2}\rho^{a^{\dagger},a}(\vec{n})\nonumber\\
&  +\sum_{\vec{n},l,S}\varepsilon_{\triangledown_{l}^{2}\rho_{S}}(\vec
{n},n_{t})\triangledown_{l}^{2}\rho_{S}^{a^{\dagger},a}(\vec{n})+\sum_{\vec
{n},l}\varepsilon_{\Pi_{l}}(\vec{n},n_{t})\Pi_{l}^{a^{\dagger},a}(\vec
{n})+\sum_{\vec{n},l,S}\varepsilon_{\Pi_{l,S}}(\vec{n},n_{t})\Pi
_{l,S}^{a^{\dagger},a}(\vec{n}),
\end{align}
the isospin-dependent couplings are%
\begin{align}
U_{I^{2}}^{(n_{t})}(\varepsilon)  &  =\sum_{\vec{n},I}\varepsilon_{\rho_{I}%
}(\vec{n},n_{t})\rho_{I}^{a^{\dagger},a}(\vec{n})+\sum_{\vec{n},S,I}%
\varepsilon_{\rho_{S,I}}(\vec{n},n_{t})\rho_{S,I}^{a^{\dagger},a}(\vec
{n})+\sum_{\vec{n},S,I}\varepsilon_{\Delta_{S}\rho_{I}}(\vec{n},n_{t}%
)\Delta_{S}\rho_{I}^{a^{\dagger},a}(\vec{n})\nonumber\\
&  +\sum_{\vec{n},S,S^{\prime},I}\varepsilon_{\Delta_{S}\rho_{S^{\prime},I}%
}(\vec{n},n_{t})\Delta_{S}\rho_{S^{\prime},I}^{a^{\dagger},a}(\vec{n}%
)+\sum_{\vec{n},l,I}\varepsilon_{\triangledown_{l}^{2}\rho_{I}}(\vec{n}%
,n_{t})\triangledown_{l}^{2}\rho_{I}^{a^{\dagger},a}(\vec{n})\nonumber\\
&  +\sum_{\vec{n},l,S,I}\varepsilon_{\triangledown_{l}^{2}\rho_{S,I}}(\vec
{n},n_{t})\triangledown_{l}^{2}\rho_{S,I}^{a^{\dagger},a}(\vec{n})+\sum
_{\vec{n},l,I}\varepsilon_{\Pi_{l,I}}(\vec{n},n_{t})\Pi_{l,I}^{a^{\dagger}%
,a}(\vec{n})+\sum_{\vec{n},l,S,I}\varepsilon_{\Pi_{l,S,I}}(\vec{n},n_{t}%
)\Pi_{l,S,I}^{a^{\dagger},a}(\vec{n}).
\end{align}
All of the NLO, NNLO, isospin-breaking, and Coulomb interactions are generated
by functional derivatives with respect to the external source fields.

We extract the properties of the ground state using Euclidean-time projection.
\ Let $\left\vert \Psi^{\text{free}}\right\rangle $ be a Slater determinant of
free-particle standing waves in a periodic cube for some chosen number of
nucleons and quantum numbers. \ Let $M_{\text{SU(4)}\not \pi }^{(n_{t})}$ be
an auxiliary-field transfer matrix at time step $n_{t}$,%
\begin{equation}
M_{\text{SU(4)}\not \pi }^{(n_{t})}(s)=\colon\exp\left[  -H_{\text{free}%
}\alpha_{t}+\sqrt{-C_{\text{SU(4)}\not \pi }\alpha_{t}}\sum_{\vec{n}}s(\vec
{n},n_{t})\rho^{a^{\dag},a}(\vec{n})\right]  :.
\end{equation}
We use the operator $M_{\text{SU(4)}\not \pi }^{(n_{t})}(s)$ to set up the
initial state for the lattice calculation,%
\begin{equation}
\left\vert \Psi(t^{\prime})\right\rangle =\left(  M_{\text{SU(4)}\not \pi
}\right)  ^{L_{t_{o}}}\left\vert \Psi^{\text{free}}\right\rangle ,
\label{L_t_o}%
\end{equation}
where $t^{\prime}=L_{t_{o}}\alpha_{t}$ and $L_{t_{o}}$ is the number of
\textquotedblleft outer\textquotedblright\ time steps. \ As the notation
suggests, the operator $M_{\text{SU(4)}\not \pi }^{(n_{t})}(s)$ is invariant
under Wigner's SU(4) symmetry \cite{Wigner:1937}. \ The repeated
multiplication by $M_{\text{SU(4)}\not \pi }^{(n_{t})}(s)$ acts as an
approximate low-energy filter. \ This part of the Euclidean-time propagation
is positive definite for any even number of nucleons invariant under the SU(4)
symmetry \cite{Lee:2004hc,Chen:2004rq,Lee:2007eu}.

The Euclidean-time amplitude $Z(t)$ is defined as%
\begin{equation}
Z(t)=\left\langle \Psi(t^{\prime})\right\vert \left(  M_{\text{LO}}\right)
^{L_{t_{i}}}\left\vert \Psi(t^{\prime})\right\rangle , \label{L_t_i}%
\end{equation}
where $t=L_{t_{i}}\alpha_{t}$ and $L_{t_{i}}$ is the number of
\textquotedblleft inner\textquotedblright\ time steps. \ The transient energy
at time $t+\alpha_{t}/2$ is calculated by taking a numerical derivative of the
logarithm of $Z(t)$,%
\begin{equation}
e^{-E_{\text{LO}}(t+\alpha_{t}/2)\cdot\alpha_{t}}=\frac{Z(t+\alpha_{t})}%
{Z(t)}.
\end{equation}
The ground state energy $E_{0,\text{LO}}$ equals the asymptotic limit of the
transient energy,%
\begin{equation}
E_{0,\text{LO}}=\lim_{t\rightarrow\infty}E_{\text{LO}}(t+\alpha_{t}/2).
\end{equation}

We calculate Euclidean-time projection amplitudes using the auxiliary-field
formalism. \ For a given configuration of auxiliary and pion fields, the
contribution to the amplitude $Z(t)$ is proportional to the determinant of an
$A\times A$ matrix of one-body amplitudes, where $A$ is the number of
nucleons. \ Integrations over auxiliary and pion field configurations are
computed using hybrid Monte Carlo. \ Details of the method can be found in
Ref.~\cite{Lee:2005fk,Lee:2006hr,Borasoy:2006qn,Lee:2008fa}.

The perturbative contributions from NLO, NNLO, isospin-breaking, and Coulomb
interactions are computed order-by-order in perturbation theory. \ For the
first-order perturbative correction to the energy, it suffices to compute
operator expectation values. \ For general operator $O$ we define the
Euclidean-time amplitude,%
\begin{equation}
Z_{O}(t)=\left\langle \Psi(t^{\prime})\right\vert \left(  M_{\text{LO}%
}\right)  ^{L_{t_{i}}/2}O\,\left(  M_{\text{LO}}\right)  ^{L_{t_{i}}%
/2}\left\vert \Psi(t^{\prime})\right\rangle .
\end{equation}
The expectation value of $O$ for $\left\vert \Psi_{0}\right\rangle $ is
extracted by taking the large $t$ limit of the ratio of $Z_{O}(t)$ and $Z(t)$,%
\begin{equation}
\lim_{t\rightarrow\infty}\frac{Z_{O}(t)}{Z(t)}=\left\langle \Psi
_{0}\right\vert O\left\vert \Psi_{0}\right\rangle .
\end{equation}
In the appendix we show precise numerical tests of the equivalence of the
auxiliary-field Monte Carlo formalism and the original transfer matrix formalism.

\section{Results for helium-4}

We compute the ground state energy for helium-4 in a periodic box of length
$9.9$~fm. \ For $\left\vert \Psi^{\text{free}}\right\rangle $ we take the
Slater determinant formed by standing waves,%
\begin{equation}
\left\langle 0\right\vert a_{i,j}(\vec{n})\left\vert \psi_{1}\right\rangle
\propto\delta_{i,0}\delta_{j,1},\qquad\left\langle 0\right\vert a_{i,j}%
(\vec{n})\left\vert \psi_{2}\right\rangle \propto\delta_{i,0}\delta_{j,0},
\end{equation}%
\begin{equation}
\left\langle 0\right\vert a_{i,j}(\vec{n})\left\vert \psi_{3}\right\rangle
\propto\delta_{i,1}\delta_{j,1},\qquad\left\langle 0\right\vert a_{i,j}%
(\vec{n})\left\vert \psi_{4}\right\rangle \propto\delta_{i,1}\delta
_{j,0}\text{.}%
\end{equation}
This produces a state with zero total momentum and the quantum numbers of the
helium-4 ground state. \ For each value of the Euclidean time, $t$, we use
$2048$ processors to generate about $5\times10^{6}$ hybrid Monte Carlo
trajectories. \ Each processor runs independent trajectories, and averages and
stochastic errors are calculated from the distribution of results from all processors.

For the numerical extrapolation in $t$, we use a decaying exponential for the
leading-order energy,%
\begin{equation}
E_{\text{LO}}(t)\approx E_{0,\text{LO}}+A_{\text{LO}}e^{-\delta E\cdot t}.
\label{asymptotic1}%
\end{equation}
For each of the perturbative energy corrections from NLO, isospin-breaking
(IB), electromagnetic (EM), and NNLO interactions we use%
\begin{equation}
\Delta E(t)\approx\Delta E_{0}+\Delta Ae^{-\delta E\cdot t/2}.
\label{asymptotic2}%
\end{equation}
The unknown parameters $E_{0,\text{LO}}$, $A_{\text{LO}}$, $\Delta A$, and
$\delta E$, are determined by least squares fitting. \ The $e^{-\delta E\cdot
t}$ dependence in Eq.~(\ref{asymptotic1}) gives the contribution of low-energy
excitations with energy gap $\delta E$ above the ground state. \ The
$e^{-\delta E\cdot t/2}$ dependence in Eq.~(\ref{asymptotic2}) gives the
contribution of matrix elements between the ground state and excitations at
energy gap $\delta E$.

Given the finite interval over which we measure the Euclidean-time dependence,
we expect some exponential dependence from other energy excitations not at
energy $\delta E$ above the ground state. \ In order to estimate the size of
the induced systematic errors, we generate an ensemble of different
exponential fits which include dropping the two first two data points and then
dropping the last two data points. \ This gives some estimate of the spread in
energies of contributing higher energy states. \ In the following we quote
total extrapolation errors which include the uncertainty due to the stochastic
errors and the effect of the distribution in $\delta E$. \ In future studies
we hope to improve this process further by considering different initial
states in order to triangulate a common extrapolated value at infinite $t$.

\begin{figure}[ptb]%
\centering
\includegraphics[
height=2.5235in,
width=4.1269in
]%
{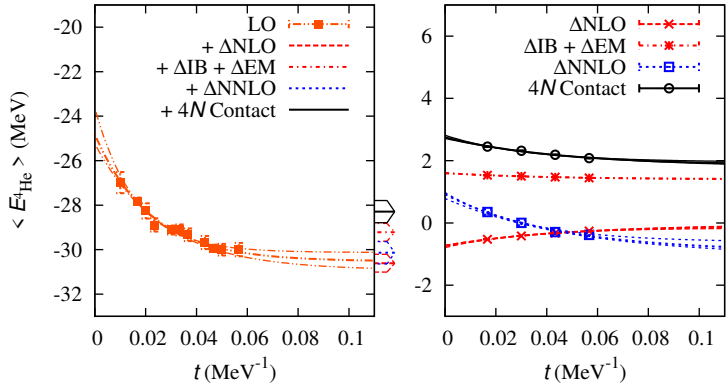}%
\caption{Ground state energy for helium-4 as a function of Euclidean time
projection. \ See text for details.}%
\label{alpha}%
\end{figure}

In Fig.~\ref{alpha} we show the energy versus Euclidean time projection for
the helium-4 ground state with LO, NLO, IB, EM, and NNLO interactions. \ The
plot on the left shows the leading-order results and the extrapolated
$t\rightarrow\infty$ values for the higher-order contributions added
cumulatively. \ These cumulative results are shown with error bars on the
right edge of the plot. \ The plot on the right shows the higher-order
corrections separately. \ For each case we show the best fit as well as the
one standard-deviation bound. \ We estimate this bound by generating an
ensemble of fits determined with added random Gaussian noise proportional to
the error bars of each data point and also varying the number of fitted data
points. \ These results are similar to those found in
Ref.~\cite{Epelbaum:2009zs} using the LO$_{2}$ action. \ For $c_{D}=1$ we get
$-30.5(4)$~MeV at LO, $-30.6(4)$~MeV at NLO, $-29.2(4)$~MeV at NLO with IB and
EM corrections, and $-30.1(5)$~MeV at NNLO. \ When the bare interaction
$D_{\text{effective}}^{(4N)}$ is held fixed, the helium-4 energy decreases
$0.4(1)$~MeV for each unit increase in $c_{D}$.

Apart from direct comparisons with experimental data, an independent estimate
of systematic errors due to truncation of higher-order terms can be made by
comparing the differences among the lattice results at each order, LO, NLO,
and NNLO. \ One caveat here is that sometimes the differences can be unusually
small, either by chance or due to underlying physics. \ For example there is
only a very small difference between the LO and NLO energies for helium-4.
\ This can be explained by the fact that the interactions for helium-4 are
predominantly in the $S$-channels, and the improved LO$_{3}$ action is already
quite accurate for $S$-wave scattering. \ For helium-4 we estimate a residual
error of size about $1$ MeV for the omitted interactions. \ This appears
consistent with the $1.8$ MeV deviation between the NNLO result and the
physical binding energy for helium-4.

For nuclei beyond $A=4$, we will test the universality hypothesis for
higher-order interactions by tuning the effective four-nucleon contact
interaction $D_{\text{effective}}^{\prime(4N)}$ to give the physical helium-4
energy of $-28.3$~MeV. \ The contribution of the effective four-nucleon
contact interaction to the helium-4 energy is shown in Fig.~\ref{alpha}.%

\section{Results for lithium-6}

We compute the ground state energy for lithium-6 in a periodic box of length
$9.9$~fm. \ For $\left\vert \Psi^{\text{free}}\right\rangle $ we choose
standing waves,%
\begin{equation}
\left\langle 0\right\vert a_{i,j}(\vec{n})\left\vert \psi_{1}\right\rangle
\propto\delta_{i,1}\delta_{j,1},\qquad\left\langle 0\right\vert a_{i,j}%
(\vec{n})\left\vert \psi_{2}\right\rangle \propto\delta_{i,1}\delta
_{j,0}\text{.}%
\end{equation}%
\begin{equation}
\left\langle 0\right\vert a_{i,j}(\vec{n})\left\vert \psi_{3}\right\rangle
\propto\delta_{i,0}\delta_{j,1}\cos\tfrac{2\pi n_{3}}{L},\qquad\left\langle
0\right\vert a_{i,j}(\vec{n})\left\vert \psi_{4}\right\rangle \propto
\delta_{i,0}\delta_{j,0}\cos\tfrac{2\pi n_{3}}{L},
\end{equation}%
\begin{equation}
\left\langle 0\right\vert a_{i,j}(\vec{n})\left\vert \psi_{5}\right\rangle
\propto\delta_{i,0}\delta_{j,1}\sin\tfrac{2\pi n_{3}}{L},\qquad\left\langle
0\right\vert a_{i,j}(\vec{n})\left\vert \psi_{6}\right\rangle \propto
\delta_{i,0}\delta_{j,0}\sin\tfrac{2\pi n_{3}}{L}.
\end{equation}
This combination produces a state with zero total momentum and the quantum
numbers of the lithium-6 ground state. \ For each value of $t$ a total of
about $5\times10^{6}$ hybrid Monte Carlo trajectories are generated by $2048$ processors.

\begin{figure}[ptb]%
\centering
\includegraphics[
height=2.5235in,
width=4.1269in
]%
{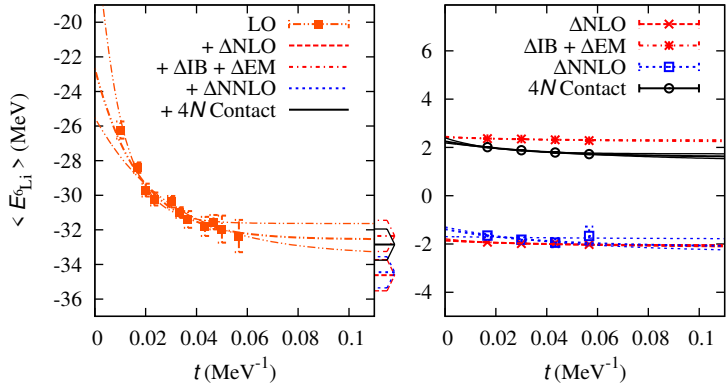}%
\caption{Ground state energy for lithium-6 as a function of Euclidean time
projection. \ See text for details.}%
\label{lithium6}%
\end{figure}

In Fig.~\ref{lithium6} we show the energy versus Euclidean time projection for
lithium-6. \ For the numerical extrapolation in $t$ we use the same decaying
exponential functions in Eq.~(\ref{asymptotic1}-\ref{asymptotic2}). \ We show
the best fit as well as the one standard-deviation bound. \ For $c_{D}=1$ we
get $-32.6(9)$~MeV at LO, $-34.6(9)$~MeV at NLO, $-32.4(9)$~MeV at NLO with IB
and EM corrections, and $-34.5(9)$~MeV at NNLO. $\ $Our error estimate due to
truncation at NNLO\ is about $2$~MeV. \ Adding the contribution of the
effective four-nucleon interaction $D_{\text{effective}}^{\prime(4N)}$ to the
NNLO result gives $-32.9(9)$~MeV. \ This lies within error bars of the
physical value $-32.0$~MeV. \ However we expect some overbinding due to the
finite periodic volume. \ The finite volume analysis in
Ref.~\cite{Epelbaum:2009zs} found a finite volume dependence of less than
$1$~MeV for the helium-4 ground state in a periodic box of length $9.9$~fm.
\ However a larger effect is expected for lithium-6 due to the larger
spatial\ distribution of the two $P$-shell nucleons. \ Further calculations at
varying volumes will be needed to determine this volume dependence.

Compared with helium-4, there is a much larger difference between the LO and
NLO energies for lithium-6. \ This may indicate additional binding coming from
the NLO corrections in $P$-wave channels. \ The dependence of the energy on
$c_{D}$ can be analyzed in several different ways. \ When the bare interaction
$D_{\text{effective}}^{(4N)}$ is held fixed, the lithium-6 energy decreases
$0.7(1)$~MeV for each unit increase in $c_{D}$. \ When the effective
four-nucleon interaction is adjusted according to the physical helium-4
energy, the lithium-6 energy decreases $0.35(5)$~MeV per unit increase in
$c_{D}$.%

\section{Results for carbon-12}

We compute the ground state energy of carbon-12 in a periodic box of length
$13.8$~fm. \ For $\left\vert \Psi^{\text{free}}\right\rangle $ we take the
Slater determinant formed by standing waves,%
\begin{equation}
\left\langle 0\right\vert a_{i,j}(\vec{n})\left\vert \psi_{4k+1}\right\rangle
\propto\delta_{i,0}\delta_{j,1}f_{k}(\vec{n}),\qquad\left\langle 0\right\vert
a_{i,j}(\vec{n})\left\vert \psi_{4k+2}\right\rangle \propto\delta_{i,0}%
\delta_{j,0}f_{k}(\vec{n}),
\end{equation}%
\begin{equation}
\left\langle 0\right\vert a_{i,j}(\vec{n})\left\vert \psi_{4k+3}\right\rangle
\propto\delta_{i,1}\delta_{j,1}f_{k}(\vec{n}),\qquad\left\langle 0\right\vert
a_{i,j}(\vec{n})\left\vert \psi_{4k+4}\right\rangle \propto\delta_{i,1}%
\delta_{j,0}f_{k}(\vec{n})\text{,}%
\end{equation}
where%
\begin{equation}
f_{0}(\vec{n})=1,\quad f_{1}(\vec{n})=\cos\tfrac{2\pi n_{3}}{L},\quad
f_{2}(\vec{n})=\sin\tfrac{2\pi n_{3}}{L}.
\end{equation}
This combination produces a state with zero total momentum and the quantum
numbers of the carbon-12 ground state. \ For each value of $t$ a total of
$2\times10^{6}$ hybrid Monte Carlo trajectories are generated by $2048$
processors. \ 

\begin{figure}[ptb]%
\centering
\includegraphics[
height=2.5244in,
width=4.1277in
]%
{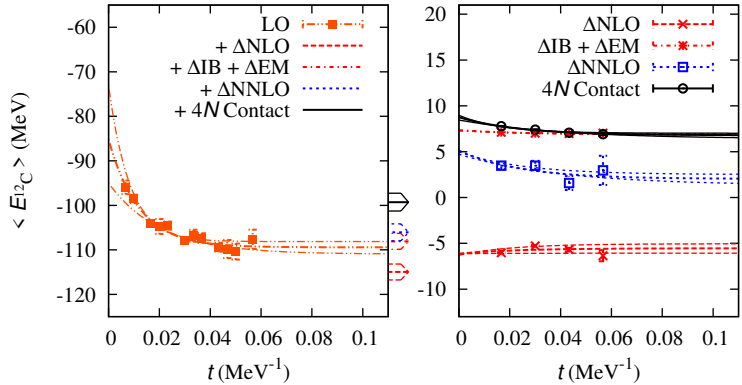}%
\caption{Ground state energy for carbon-12 as a function of Euclidean time
projection. \ See text for details.}%
\label{carbon12}%
\end{figure}

Fig.~\ref{carbon12} shows the energy versus Euclidean time projection for
carbon-12. \ For $c_{D}=1$ we get $-109(2)$~MeV at LO, $-115(2)$~MeV at NLO,
$-108(2)$~MeV at NLO with IB and EM corrections, and $-106(2)$~MeV at NNLO.
$\ $Our error estimate due to truncation at NNLO\ is about $5$~MeV. \ The
small $2$~MeV difference between NLO and NNLO results is due to a cancellation
of several larger contributions. Adding the contribution of the effective
four-nucleon interaction $D_{\text{effective}}^{\prime(4N)}$ to the NNLO
result gives $-99(2)$~MeV. \ This is an overbinding of $7\%$ compared to the
physical value, $-92.2$~MeV. \ While this agreement as a final result would
not be bad, an overbinding of $7\%$ is actually a reasonable estimate of the
finite volume correction for carbon-12 in a periodic box of length $13.8$~fm.
\ If so the error at infinite volume would in fact be much smaller than $7\%$.
\ Further calculations at varying volumes will be needed to measure the volume dependence.

When the bare interaction $D_{\text{effective}}^{(4N)}$ is held fixed, the
carbon-12 energy decreases $1.7(3)$~MeV per unit increase in $c_{D}$.\ \ When
the effective four-nucleon interaction is adjusted according to the physical
helium-4 energy, the carbon-12 energy decreases only $0.3(1)$~MeV per unit
increase in $c_{D}$. \ The much reduced dependence upon on $c_{D}$ is
consistent with our universality hypothesis regarding systematic errors. \ In
three-nucleon systems the value of $c_{D}$ is difficult to resolve due to
similarities of the one-pion exchange three-nucleon interaction and the
three-nucleon contact interaction at low energies. \ For systems with four or
more nucleons, the difference between these three-nucleon interactions becomes
significant. \ However our universality hypothesis suggests that this
difference behaves like an effective four-nucleon contact interaction. \ This
explains why the dependence on $c_{D}$ goes away when we include an effective
four-nucleon contact interaction tuned to the physical helium-4 energy.%

\section{Summary and comments}

In this paper we have presented several new methods and results in lattice
effective field theory. \ We described the first lattice results for lithium-6
and carbon-12 using chiral effective field theory. \ This represents a
significant advance in the range of problems accessible using lattice
effective field theory. \ We also detailed the first lattice calculations to
include isospin-breaking and Coulomb interactions, and computed the energy
splitting between helium-3 and the triton. \ The accuracy of the lattice
calculations presented here are competitive with recent calculations obtained
using other ab initio methods. \ Coupled cluster calculations without
three-nucleon interactions are accurate to within $1$~MeV per nucleon for
medium mass nuclei \cite{Hagen:2007hi}. \ \ Constrained-path Green's function
Monte Carlo calculations generally have an accuracy of $1\%-2\%$ in energy for
nuclei $A\leq12$. \ The most recent result for carbon-12 is $-93.2(6)$~MeV
using AV18 and the IL7 three-nucleon force \cite{Pieper:2009aa}. \ The most
recent no-core shell model calculation for carbon-12 with the JISP16 $NN$
interaction considers two different extrapolation methods to obtain values
$-93.9(1.1)$~MeV and $-95.1(2.7)$~MeV \cite{Maris:2008ax}. \ We also mention
some recent lattice QCD simulations in the strong coupling limit. \ While
quite different from physical nuclei, the strong coupling analog of nuclei
have been simulated for up to twelve nucleons \cite{deForcrand:2009dh}.

Future lattice studies should look at probing large volumes, decreasing the
lattice spacing, and including higher-order interactions. \ The computational
scaling with the number of nucleons suggests that larger nuclei are also
possible. \ At fixed volume we find that the time required by one processor to
generate one HMC\ trajectory scales with the number of nucleons as $A^{1.7}$
for $A\leq16$. \ For carbon-12 calculations the time required by one processor
to generate one HMC\ trajectory scales with volume as $V^{1.5}$. \ For nuclei
with $S=0$ and $I=0$ the average sign $\left\langle e^{i\theta}\right\rangle $
scales as $e^{-0.11A}$. \ From this scaling data we estimate that a simulation
of oxygen-16 would require about $1.8$ TFlop-yr.

Lattice effective field theory should prove a useful tool for few-body
calculations of nuclei as well as many-body calculations of neutron and
nuclear matter. \ The method is also quite attractive theoretically as it uses
only the general principles of effective field theory. All systematic errors
are introduced up front when defining the truncated low-energy effective
theory. \ This eliminates approximation errors tied with a specific
calculational tool, physical system, or observable. \ The reduction of these
errors is not necessarily easy. \ However they can be clearly identified as
either missing operators in the lattice action, finite volume effects, or
errors from finite\ Euclidean-time extrapolation. \ Future studies can then
improve upon existing calculations in a straightforward manner.

\section*{Acknowledgements}

Partial financial support from the Deutsche Forschungsgemeinschaft (SFB/TR
16), Helmholtz Association (contract number VH-NG-222 and VH-VI-231),
BMBF\ (grant 06BN9006), and U.S. Department of Energy (DE-FG02-03ER41260) are
acknowledged. \ This work was further supported by the EU HadronPhysics2
project \textquotedblleft Study of strongly interacting
matter\textquotedblright. \ The computational resources for this project were
provided by the J\"{u}lich Supercomputing Centre at the Forschungszentrum J\"{u}lich.

\appendix

\section{Lattice notation}

The vector $\vec{n}$ represents integer-valued lattice vectors on a
three-dimensional spatial lattice, and $\vec{p},$ $\vec{q},$ $\vec{k}$
represent integer-valued momentum lattice vectors.$\ \ \hat{l}=\hat{1}$,
$\hat{2}$, $\hat{3}$ are unit lattice vectors in the spatial directions, $a$
is the spatial lattice spacing, and $L$ is the length of the cubic spatial
lattice in each direction. \ The lattice time step is $a_{t}$, and $n_{t}$
labels the number of time steps. \ We define $\alpha_{t}$ as the ratio between
lattice spacings, $\alpha_{t}=a_{t}/a$. \ Throughout our lattice discussion we
use dimensionless parameters and operators, which correspond with physical
values multiplied by the appropriate power of $a$. \ Final results are
presented in physical units with the corresponding unit stated explicitly.

We use $a$ and $a^{\dagger}$ to denote annihilation and creation operators.
\ We make explicit all spin and isospin indices,%
\begin{align}
a_{0,0}  &  =a_{\uparrow,p},\text{ \ }a_{0,1}=a_{\uparrow,n},\\
a_{1,0}  &  =a_{\downarrow,p},\text{ \ }a_{1,1}=a_{\downarrow,n}.
\end{align}
The first subscript is for spin and the second subscript is for isospin. \ We
use $\tau_{I}$ with $I=1,2,3$ to represent Pauli matrices acting in isospin
space and $\sigma_{S}$ with $S=1,2,3$ to represent Pauli matrices acting in
spin space. \ For the free nucleon we use the $O(a^{4})$-improved lattice
Hamiltonian,%
\begin{align}
H_{\text{free}}  &  =\frac{49}{12m}\sum_{\vec{n}}\sum_{i,j=0,1}a_{i,j}%
^{\dagger}(\vec{n})a_{i,j}(\vec{n})\nonumber\\
&  -\frac{3}{4m}\sum_{\vec{n}}\sum_{i,j=0,1}\sum_{l=1,2,3}\left[
a_{i,j}^{\dagger}(\vec{n})a_{i,j}(\vec{n}+\hat{l})+a_{i,j}^{\dagger}(\vec
{n})a_{i,j}(\vec{n}-\hat{l})\right] \nonumber\\
&  +\frac{3}{40m}\sum_{\vec{n}}\sum_{i,j=0,1}\sum_{l=1,2,3}\left[
a_{i,j}^{\dagger}(\vec{n})a_{i,j}(\vec{n}+2\hat{l})+a_{i,j}^{\dagger}(\vec
{n})a_{i,j}(\vec{n}-2\hat{l})\right] \nonumber\\
&  -\frac{1}{180m}\sum_{\vec{n}}\sum_{i,j=0,1}\sum_{l=1,2,3}\left[
a_{i,j}^{\dagger}(\vec{n})a_{i,j}(\vec{n}+3\hat{l})+a_{i,j}^{\dagger}(\vec
{n})a_{i,j}(\vec{n}-3\hat{l})\right]  .
\end{align}

The eight vertices of a unit cube on the lattice is used to define spatial
derivatives. \ For each spatial direction $l=1,2,3$ and any lattice function
$f(\vec{n})$, let%
\begin{equation}
\Delta_{l}f(\vec{n})=\frac{1}{4}\sum_{\substack{\nu_{1},\nu_{2},\nu_{3}%
=0,1}}(-1)^{\nu_{l}+1}f(\vec{n}+\vec{\nu}),\qquad\vec{\nu}=\nu_{1}\hat{1}%
+\nu_{2}\hat{2}+\nu_{3}\hat{3}. \label{derivative}%
\end{equation}
We also define the double spatial derivative along direction $l$,%
\begin{equation}
\triangledown_{l}^{2}f(\vec{n})=f(\vec{n}+\hat{l})+f(\vec{n}-\hat{l}%
)-2f(\vec{n}).
\end{equation}
For the three-body NNLO interactions we also use the notation%
\begin{equation}
\square f(\vec{n})=\frac{1}{8}\sum_{\substack{\nu_{1},\nu_{2},\nu_{3}%
=0,1}}f(\vec{n}+\vec{\nu}),\qquad\vec{\nu}=\nu_{1}\hat{1}+\nu_{2}\hat{2}%
+\nu_{3}\hat{3}.
\end{equation}

\subsection{Local densities and currents}

We define the local density,%
\begin{equation}
\rho^{a^{\dagger},a}(\vec{n})=\sum_{i,j=0,1}a_{i,j}^{\dagger}(\vec{n}%
)a_{i,j}(\vec{n}),
\end{equation}
which is invariant under Wigner's SU(4) symmetry \cite{Wigner:1937}.
\ Similarly we define the local spin density for $S=1,2,3,$%
\begin{equation}
\rho_{S}^{a^{\dagger},a}(\vec{n})=\sum_{i,j,i^{\prime}=0,1}a_{i,j}^{\dagger
}(\vec{n})\left[  \sigma_{S}\right]  _{ii^{\prime}}a_{i^{\prime},j}(\vec{n}),
\end{equation}
isospin density for $I=1,2,3,$%
\begin{equation}
\rho_{I}^{a^{\dagger},a}(\vec{n})=\sum_{i,j,j^{\prime}=0,1}a_{i,j}^{\dagger
}(\vec{n})\left[  \tau_{I}\right]  _{jj^{\prime}}a_{i,j^{\prime}}(\vec{n}),
\end{equation}
and spin-isospin density for $S,I=1,2,3,$%
\begin{equation}
\rho_{S,I}^{a^{\dagger},a}(\vec{n})=\sum_{i,j,i^{\prime},j^{\prime}%
=0,1}a_{i,j}^{\dagger}(\vec{n})\left[  \sigma_{S}\right]  _{ii^{\prime}%
}\left[  \tau_{I}\right]  _{jj^{\prime}}a_{i^{\prime},j^{\prime}}(\vec{n}).
\end{equation}

For each static density we also have an associated current density. \ Similar
to the definition of the lattice derivative $\Delta_{l}$ in
Eq.~(\ref{derivative}), we use the eight vertices of a unit cube,
\begin{equation}
\vec{\nu}=\nu_{1}\hat{1}+\nu_{2}\hat{2}+\nu_{3}\hat{3},
\end{equation}
for $\nu_{1},\nu_{2},\nu_{3}=0,1$. \ Let $\vec{\nu}(-l)$ for $l=1,2,3$ be the
result of reflecting the $l^{\text{th}}$-component of $\vec{\nu}$ about the
center of the cube,%
\begin{equation}
\vec{\nu}(-l)=\vec{\nu}+(1-2\nu_{l})\hat{l}.
\end{equation}
Omitting factors of $i$ and $1/m$, we can write the $l^{\text{th}}$-component
of the SU(4)-invariant current density as%
\begin{equation}
\Pi_{l}^{a^{\dagger},a}(\vec{n})=\frac{1}{4}\sum_{\substack{\nu_{1},\nu
_{2},\nu_{3}=0,1}}\sum_{i,j=0,1}(-1)^{\nu_{l}+1}a_{i,j}^{\dagger}(\vec{n}%
+\vec{\nu}(-l))a_{i,j}(\vec{n}+\vec{\nu}).
\end{equation}
Similarly the $l^{\text{th}}$-component of spin current density is%
\begin{equation}
\Pi_{l,S}^{a^{\dagger},a}(\vec{n})=\frac{1}{4}\sum_{\substack{\nu_{1},\nu
_{2},\nu_{3}=0,1}}\sum_{i,j,i^{\prime}=0,1}(-1)^{\nu_{l}+1}a_{i,j}^{\dagger
}(\vec{n}+\vec{\nu}(-l))\left[  \sigma_{S}\right]  _{ii^{\prime}}a_{i^{\prime
},j}(\vec{n}+\vec{\nu}),
\end{equation}
$l^{\text{th}}$-component of isospin current density is%
\begin{equation}
\Pi_{l,I}^{a^{\dagger},a}(\vec{n})=\frac{1}{4}\sum_{\substack{\nu_{1},\nu
_{2},\nu_{3}=0,1}}\sum_{i,j,j^{\prime}=0,1}(-1)^{\nu_{l}+1}a_{i,j}^{\dagger
}(\vec{n}+\vec{\nu}(-l))\left[  \tau_{I}\right]  _{jj^{\prime}}a_{i,j^{\prime
}}(\vec{n}+\vec{\nu}),
\end{equation}
and $l^{\text{th}}$-component of spin-isospin current density is%
\begin{equation}
\Pi_{l,S,I}^{a^{\dagger},a}(\vec{n})=\frac{1}{4}\sum_{\substack{\nu_{1}%
,\nu_{2},\nu_{3}=0,1}}\sum_{i,j,i^{\prime},j^{\prime}=0,1}(-1)^{\nu_{l}%
+1}a_{i,j}^{\dagger}(\vec{n}+\vec{\nu}(-l))\left[  \sigma_{S}\right]
_{ii^{\prime}}\left[  \tau_{I}\right]  _{jj^{\prime}}a_{i^{\prime},j^{\prime}%
}(\vec{n}+\vec{\nu}).
\end{equation}

\subsection{Instantaneous free pion action}

The lattice action for free pions with purely instantaneous propagation is%
\begin{equation}
S_{\pi\pi}(\pi_{I})=\alpha_{t}(\tfrac{m_{\pi}^{2}}{2}+3)\sum_{\vec{n},n_{t}%
,I}\pi_{I}(\vec{n},n_{t})\pi_{I}(\vec{n},n_{t})-\alpha_{t}\sum_{\vec{n}%
,n_{t},I,l}\pi_{I}(\vec{n},n_{t})\pi_{I}(\vec{n}+\hat{l},n_{t}),
\end{equation}
where $\pi_{I}$ is the pion field labelled with isospin index $I$, and
$m_{\pi}=m_{\pi^{0}}$. \ It is convenient to define a rescaled pion field,
$\pi_{I}^{\prime}$,%
\begin{equation}
\pi_{I}^{\prime}(\vec{n},n_{t})=\sqrt{q_{\pi}}\pi_{I}(\vec{n},n_{t}),
\end{equation}%
\begin{equation}
q_{\pi}=\alpha_{t}(m_{\pi}^{2}+6).
\end{equation}
Then%
\begin{equation}
S_{\pi\pi}(\pi_{I}^{\prime})=\frac{1}{2}\sum_{\vec{n},n_{t},I}\pi_{I}^{\prime
}(\vec{n},n_{t})\pi_{I}^{\prime}(\vec{n},n_{t})-\frac{\alpha_{t}}{q_{\pi}}%
\sum_{\vec{n},n_{t},I,l}\pi_{I}^{\prime}(\vec{n},n_{t})\pi_{I}^{\prime}%
(\vec{n}+\hat{l},n_{t}). \label{pionaction}%
\end{equation}

In momentum space the action is%
\begin{equation}
S_{\pi\pi}(\pi_{I}^{\prime})=\frac{1}{L^{3}}\sum_{I,\vec{k}}\pi_{I}^{\prime
}(-\vec{k},n_{t})\pi_{I}^{\prime}(\vec{k},n_{t})\left[  \frac{1}{2}%
-\frac{\alpha_{t}}{q_{\pi}}\sum_{l}\cos k_{l}\right]  .
\end{equation}
The instantaneous pion correlation function at spatial separation $\vec{n}$ is%
\begin{align}
\left\langle \pi_{I}^{\prime}(\vec{n},n_{t})\pi_{I}^{\prime}(\vec{0}%
,n_{t})\right\rangle  &  =\frac{\int D\pi_{I}^{\prime}\;\pi_{I}^{\prime}%
(\vec{n},n_{t})\pi_{I}^{\prime}(\vec{0},n_{t})\;\exp\left[  -S_{\pi\pi
}\right]  }{\int D\pi_{I}^{\prime}\;\exp\left[  -S_{\pi\pi}\right]  }\text{
\ (no sum on }I\text{)}\nonumber\\
&  =\frac{1}{L^{3}}\sum_{\vec{k}}e^{-i\vec{k}\cdot\vec{n}}D_{\pi}(\vec{k}),
\end{align}
where%
\begin{equation}
D_{\pi}(\vec{k})=\frac{1}{1-\tfrac{2\alpha_{t}}{q_{\pi}}\sum_{l}\cos k_{l}}.
\end{equation}
It is also useful to define the two-derivative pion correlator, $G_{S_{1}%
S_{2}}(\vec{n})$,%
\begin{align}
G_{S_{1}S_{2}}(\vec{n})  &  =\left\langle \Delta_{S_{1}}\pi_{I}^{\prime}%
(\vec{n},n_{t})\Delta_{S_{2}}\pi_{I}^{\prime}(\vec{0},n_{t})\right\rangle
\text{ \ (no sum on }I\text{)}\nonumber\\
&  =\frac{1}{16}\sum_{\nu_{1},\nu_{2},\nu_{3}=0,1}\sum_{\nu_{1}^{\prime}%
,\nu_{2}^{\prime},\nu_{3}^{\prime}=0,1}(-1)^{\nu_{S_{1}}}(-1)^{\nu_{S_{2}%
}^{\prime}}\left\langle \pi_{I}^{\prime}(\vec{n}+\vec{\nu}-\vec{\nu}^{\prime
},n_{t})\pi_{I}^{\prime}(\vec{0},n_{t})\right\rangle . \label{G_SS}%
\end{align}

\subsection{Pion mass differences}

We outline the modifications that result from different masses for the charged
pion and neutral pion. \ Let%
\begin{equation}
q_{\pi}(m_{\pi^{\pm}})=\alpha_{t}(m_{\pi^{\pm}}^{2}+6),\qquad q_{\pi}%
(m_{\pi^{0}})=\alpha_{t}(m_{\pi^{0}}^{2}+6).
\end{equation}
The rescaled pion fields are then%
\begin{equation}
\pi_{1,2}^{\prime}(\vec{n},n_{t})=\sqrt{q_{\pi}(m_{\pi^{\pm}})}\pi_{1,2}%
(\vec{n},n_{t}),\qquad\pi_{3}^{\prime}(\vec{n},n_{t})=\sqrt{q_{\pi}(m_{\pi
^{0}})}\pi_{3}(\vec{n},n_{t}).
\end{equation}
The momentum-space correlators for the charged and neutral pions are%
\begin{equation}
D_{\pi}(\vec{k},m_{\pi^{\pm}})=\frac{1}{1-\tfrac{2\alpha_{t}}{q_{\pi}%
(m_{\pi^{\pm}})}\sum_{l}\cos k_{l}},
\end{equation}%
\begin{equation}
D_{\pi}(\vec{k},m_{\pi^{0}})=\frac{1}{1-\tfrac{2\alpha_{t}}{q_{\pi}(m_{\pi
^{0}})}\sum_{l}\cos k_{l}}.
\end{equation}
We can now repeat the steps in Eq.~(\ref{G_SS}) to define the two-derivative
pion correlators $G_{S_{1}S_{2}}(\vec{n},m_{\pi^{\pm}})$ and $G_{S_{1}S_{2}%
}(\vec{n},m_{\pi^{0}})$.

\section{Precision tests}

We use the three-nucleon system as a precision test of the lattice formalism
and computer codes. \ The same observables are calculated using both
auxiliary-field Monte Carlo and the exact transfer matrix without auxiliary
fields. \ We choose a small system so that stochastic errors are small enough
to expose disagreement at the $0.1\%-1\%$ level. \ We choose the spatial
length of the lattice to be $L=3$ lattice units and set the outer time steps
$L_{t_{o}}=0$ and inner time steps $L_{t_{i}}=4$. \ With $2048$ processors we
generate a total of about $10^{7}$ hybrid Monte Carlo trajectories. \ Each
processor runs completely independent trajectories, and we compute averages
and stochastic errors by comparing the results of all processors.

We choose $\left\vert \Psi^{\text{free}}\right\rangle $ to be a Slater
determinant of free-particle standing waves where%
\begin{equation}
\left\langle 0\right\vert a_{i,j}(\vec{n})\left\vert \psi_{1}\right\rangle
\propto\delta_{i,0}\delta_{j,0},\qquad\left\langle 0\right\vert a_{i,j}%
(\vec{n})\left\vert \psi_{2}\right\rangle \propto\delta_{i,1}\delta
_{j,0},\qquad\left\langle 0\right\vert a_{i,j}(\vec{n})\left\vert \psi
_{3}\right\rangle \propto\delta_{i,0}\delta_{j,1}\text{.}%
\end{equation}
The quantum numbers of this state correspond with helium-3 at zero momentum.
\ At leading order we find an energy of $-49.72(6)$~MeV for the Monte Carlo
calculation and $-49.7515$~MeV for the exact transfer matrix. \ In
Table~\ref{precision_nlo} we compare Monte Carlo results (MC) and exact
transfer matrix calculations (Exact) for the derivative of the energy with
respect to each NLO coefficient. \ Table~\ref{precision_isospin_breaking}
shows the energy shifts due to the proton-proton contact interaction and the
Coulomb interaction, and Table~\ref{precision_nnlo} shows the derivative of
the energy with respect to each NNLO coefficient. \ The numbers in parentheses
are the estimated stochastic errors. \ In all cases the agreement between
Monte Carlo results and exact transfer calculations is consistent with
estimated stochastic errors.\begin{table}[tb]
\caption{Monte Carlo results versus exact transfer matrix calculations for the
derivative of the energy with respect to NLO coefficients.}%
\label{precision_nlo}
\begin{tabular}
[c]{||c|c|c||}\hline\hline
NLO energy derivatives & MC & Exact\\\hline
$\frac{\partial\left(  \Delta E_{\text{NLO}}(t)\right)  }{\partial\left(
\Delta C\right)  }$ [$10^{4}$ MeV$^{3}$] & $3.722(3)$ & $3.72347$\\\hline
$\frac{\partial\left(  \Delta E_{\text{NLO}}(t)\right)  }{\partial\left(
\Delta C_{I^{2}}\right)  }$ [$10^{4}$ MeV$^{3}$] & $-4.530(6)$ &
$-4.53590$\\\hline
$\frac{\partial\left(  \Delta E_{\text{NLO}}(t)\right)  }{\partial\left(
C_{q^{2}}\right)  }$ [$10^{9}$ MeV$^{5}$] & $-2.055(2)$ & $-2.05383$\\\hline
$\frac{\partial\left(  \Delta E_{\text{NLO}}(t)\right)  }{\partial\left(
C_{I^{2},q^{2}}\right)  }$ [$10^{9}$ MeV$^{5}$] & $3.052(3)$ & $3.05148$%
\\\hline
$\frac{\partial\left(  \Delta E_{\text{NLO}}(t)\right)  }{\partial\left(
C_{S^{2},q^{2}}\right)  }$ [$10^{9}$ MeV$^{5}$] & $0.161(3)$ & $0.16376$%
\\\hline
$\frac{\partial\left(  \Delta E_{\text{NLO}}(t)\right)  }{\partial\left(
C_{S^{2},I^{2},q^{2}}\right)  }$ [$10^{9}$ MeV$^{5}$] & $5.240(5)$ &
$5.24260$\\\hline
$\frac{\partial\left(  \Delta E_{\text{NLO}_{3}}(t)\right)  }{\partial\left(
C_{(q\cdot S)^{2}}\right)  }$ [$10^{9}$ MeV$^{5}$] & $-1.5873(9)$ &
$-1.58896$\\\hline
$\frac{\partial\left(  \Delta E_{\text{NLO}}(t)\right)  }{\partial\left(
C_{I^{2},(q\cdot S)^{2}}\right)  }$ [$10^{9}$ MeV$^{5}$] & $6.833(3)$ &
$6.83234$\\\hline
$\frac{\partial\left(  \Delta E_{\text{NLO}}(t)\right)  }{\partial\left(
C_{(iq\times S)\cdot k}\right)  }$ [$10^{9}$ MeV$^{5}$] & $0.3356(5)$ &
$0.33702$\\\hline
$\frac{\partial\left(  \Delta E_{\text{NLO}}(t)\right)  }{\partial\left(
C_{I^{2},(iq\times S)\cdot k}\right)  }$ [$10^{9}$ MeV$^{5}$] & $-0.996(2)$ &
$-0.99656$\\\hline\hline
\end{tabular}
\end{table}

\begin{table}[tb]
\caption{Monte Carlo results versus exact transfer matrix calculations for the
energy shifts due to the proton-proton contact interaction and the Coulomb
interaction.}%
\label{precision_isospin_breaking}
\begin{tabular}
[c]{||c|c|c||}\hline\hline
IB\ and EM energy shifts & MC & Exact\\\hline
$\Delta E_{\text{pp}}(t)$ [$10^{-2}$ MeV] & 1.937(2) & 1.94128\\\hline
$\Delta E_{\text{EM}}(t)$ [$10^{-1}$ MeV] & 3.712(2) & 3.71232\\\hline\hline
\end{tabular}
\end{table}

\begin{table}[tb]
\caption{Monte Carlo results versus exact transfer matrix calculations for the
derivative of the energy with respect to NNLO coefficients.}%
\label{precision_nnlo}
\begin{tabular}
[c]{||c|c|c||}\hline\hline
NNLO energy derivatives & MC & Exact\\\hline
$\frac{\partial\left(  \Delta E_{\text{NNLO}}(t)\right)  }{\partial\left(
D_{\text{contact}}\right)  }$ [$10^{8}$ MeV$^{6}$] & $0.999(7)$ &
$1.0029$\\\hline
$\frac{\partial\left(  \Delta E_{\text{NNLO}}(t)\right)  }{\partial\left(
D_{\text{OPE}}\right)  }$ [$10^{7}$ MeV$^{5}$] & $-5.81(2)$ & $-5.8070$%
\\\hline
$\frac{\partial\left(  \Delta E_{\text{NNLO}}(t)\right)  }{\partial\left(
D_{\text{TPE1}}\right)  }$ [$10^{5}$ MeV$^{4}$] & $15.27(13)$ & $15.319$%
\\\hline
$\frac{\partial\left(  \Delta E_{\text{NNLO}}(t)\right)  }{\partial\left(
D_{\text{TPE2}}\right)  }$ [$10^{5}$ MeV$^{4}$] & $2.33(6)$ & $2.2744$\\\hline
$\frac{\partial\left(  \Delta E_{\text{NNLO}}(t)\right)  }{\partial\left(
D_{\text{TPE3}}\right)  }$ [$10^{5}$ MeV$^{4}$] & $-10.9(2)$ & $-11.032$%
\\\hline\hline
\end{tabular}
\end{table}

\bibliographystyle{apsrev}
\bibliography{References}

\begin{thebibliography}{59}
\expandafter\ifx\csname natexlab\endcsname\relax\def\natexlab#1{#1}\fi
\expandafter\ifx\csname bibnamefont\endcsname\relax
  \def\bibnamefont#1{#1}\fi
\expandafter\ifx\csname bibfnamefont\endcsname\relax
  \def\bibfnamefont#1{#1}\fi
\expandafter\ifx\csname citenamefont\endcsname\relax
  \def\citenamefont#1{#1}\fi
\expandafter\ifx\csname url\endcsname\relax
  \def\url#1{\texttt{#1}}\fi
\expandafter\ifx\csname urlprefix\endcsname\relax\def\urlprefix{URL }\fi
\providecommand{\bibinfo}[2]{#2}
\providecommand{\eprint}[2][]{\url{#2}}

\bibitem[{\citenamefont{M{\"u}ller et~al.}(2000)\citenamefont{M{\"u}ller,
  Koonin, Seki, and van Kolck}}]{Muller:1999cp}
\bibinfo{author}{\bibfnamefont{H.~M.} \bibnamefont{M{\"u}ller}},
  \bibinfo{author}{\bibfnamefont{S.~E.} \bibnamefont{Koonin}},
  \bibinfo{author}{\bibfnamefont{R.}~\bibnamefont{Seki}}, \bibnamefont{and}
  \bibinfo{author}{\bibfnamefont{U.}~\bibnamefont{van Kolck}},
  \bibinfo{journal}{Phys. Rev.} \textbf{\bibinfo{volume}{C61}},
  \bibinfo{pages}{044320} (\bibinfo{year}{2000}), \eprint{nucl-th/9910038}.

\bibitem[{\citenamefont{Lee and Sch{\"a}fer}(2005)}]{Lee:2004qd}
\bibinfo{author}{\bibfnamefont{D.}~\bibnamefont{Lee}} \bibnamefont{and}
  \bibinfo{author}{\bibfnamefont{T.}~\bibnamefont{Sch{\"a}fer}},
  \bibinfo{journal}{Phys. Rev.} \textbf{\bibinfo{volume}{C72}},
  \bibinfo{pages}{024006} (\bibinfo{year}{2005}), \eprint{nucl-th/0412002}.

\bibitem[{\citenamefont{Lee et~al.}(2004)\citenamefont{Lee, Borasoy, and
  Sch{\"a}fer}}]{Lee:2004si}
\bibinfo{author}{\bibfnamefont{D.}~\bibnamefont{Lee}},
  \bibinfo{author}{\bibfnamefont{B.}~\bibnamefont{Borasoy}}, \bibnamefont{and}
  \bibinfo{author}{\bibfnamefont{T.}~\bibnamefont{Sch{\"a}fer}},
  \bibinfo{journal}{Phys. Rev.} \textbf{\bibinfo{volume}{C70}},
  \bibinfo{pages}{014007} (\bibinfo{year}{2004}), \eprint{nucl-th/0402072}.

\bibitem[{\citenamefont{Abe and Seki}(2009)}]{Abe:2007fe}
\bibinfo{author}{\bibfnamefont{T.}~\bibnamefont{Abe}} \bibnamefont{and}
  \bibinfo{author}{\bibfnamefont{R.}~\bibnamefont{Seki}},
  \bibinfo{journal}{Phys. Rev.} \textbf{\bibinfo{volume}{C79}},
  \bibinfo{pages}{054002} (\bibinfo{year}{2009}), \eprint{arXiv:0708.2523
  [nucl-th]}.

\bibitem[{\citenamefont{Borasoy et~al.}(2008)\citenamefont{Borasoy, Epelbaum,
  Krebs, Lee, and Mei{\ss}ner}}]{Borasoy:2007vk}
\bibinfo{author}{\bibfnamefont{B.}~\bibnamefont{Borasoy}},
  \bibinfo{author}{\bibfnamefont{E.}~\bibnamefont{Epelbaum}},
  \bibinfo{author}{\bibfnamefont{H.}~\bibnamefont{Krebs}},
  \bibinfo{author}{\bibfnamefont{D.}~\bibnamefont{Lee}}, \bibnamefont{and}
  \bibinfo{author}{\bibfnamefont{U.-G.} \bibnamefont{Mei{\ss}ner}},
  \bibinfo{journal}{Eur. Phys. J.} \textbf{\bibinfo{volume}{A35}},
  \bibinfo{pages}{357} (\bibinfo{year}{2008}), \eprint{arXiv:0712.2993
  [nucl-th]}.

\bibitem[{\citenamefont{Epelbaum
  et~al.}(2009{\natexlab{a}})\citenamefont{Epelbaum, Krebs, Lee, and
  Mei{\ss}ner}}]{Epelbaum:2008vj}
\bibinfo{author}{\bibfnamefont{E.}~\bibnamefont{Epelbaum}},
  \bibinfo{author}{\bibfnamefont{H.}~\bibnamefont{Krebs}},
  \bibinfo{author}{\bibfnamefont{D.}~\bibnamefont{Lee}}, \bibnamefont{and}
  \bibinfo{author}{\bibfnamefont{U.-G.} \bibnamefont{Mei{\ss}ner}},
  \bibinfo{journal}{Eur. Phys. J.} \textbf{\bibinfo{volume}{A40}},
  \bibinfo{pages}{199} (\bibinfo{year}{2009}{\natexlab{a}}),
  \eprint{arXiv:0812.3653 [nucl-th]}.

\bibitem[{\citenamefont{Wlazlowski and Magierski}(2009)}]{Wlazlowski:2009yi}
\bibinfo{author}{\bibfnamefont{G.}~\bibnamefont{Wlazlowski}} \bibnamefont{and}
  \bibinfo{author}{\bibfnamefont{P.}~\bibnamefont{Magierski}}
  (\bibinfo{year}{2009}), \eprint{0912.0373}.

\bibitem[{\citenamefont{Borasoy et~al.}(2006)\citenamefont{Borasoy, Krebs, Lee,
  and Mei{\ss}ner}}]{Borasoy:2005yc}
\bibinfo{author}{\bibfnamefont{B.}~\bibnamefont{Borasoy}},
  \bibinfo{author}{\bibfnamefont{H.}~\bibnamefont{Krebs}},
  \bibinfo{author}{\bibfnamefont{D.}~\bibnamefont{Lee}}, \bibnamefont{and}
  \bibinfo{author}{\bibfnamefont{U.-G.} \bibnamefont{Mei{\ss}ner}},
  \bibinfo{journal}{Nucl. Phys.} \textbf{\bibinfo{volume}{A768}},
  \bibinfo{pages}{179} (\bibinfo{year}{2006}), \eprint{nucl-th/0510047}.

\bibitem[{\citenamefont{Borasoy
  et~al.}(2007{\natexlab{a}})\citenamefont{Borasoy, Epelbaum, Krebs, Lee, and
  Mei{\ss}ner}}]{Borasoy:2006qn}
\bibinfo{author}{\bibfnamefont{B.}~\bibnamefont{Borasoy}},
  \bibinfo{author}{\bibfnamefont{E.}~\bibnamefont{Epelbaum}},
  \bibinfo{author}{\bibfnamefont{H.}~\bibnamefont{Krebs}},
  \bibinfo{author}{\bibfnamefont{D.}~\bibnamefont{Lee}}, \bibnamefont{and}
  \bibinfo{author}{\bibfnamefont{U.-G.} \bibnamefont{Mei{\ss}ner}},
  \bibinfo{journal}{Eur. Phys. J.} \textbf{\bibinfo{volume}{A31}},
  \bibinfo{pages}{105} (\bibinfo{year}{2007}{\natexlab{a}}),
  \eprint{nucl-th/0611087}.

\bibitem[{\citenamefont{Epelbaum
  et~al.}(2009{\natexlab{b}})\citenamefont{Epelbaum, Krebs, Lee, and
  Mei{\ss}ner}}]{Epelbaum:2009zs}
\bibinfo{author}{\bibfnamefont{E.}~\bibnamefont{Epelbaum}},
  \bibinfo{author}{\bibfnamefont{H.}~\bibnamefont{Krebs}},
  \bibinfo{author}{\bibfnamefont{D.}~\bibnamefont{Lee}}, \bibnamefont{and}
  \bibinfo{author}{\bibfnamefont{U.~G.} \bibnamefont{Mei{\ss}ner}},
  \bibinfo{journal}{Eur. Phys. J.} \textbf{\bibinfo{volume}{A41}},
  \bibinfo{pages}{125} (\bibinfo{year}{2009}{\natexlab{b}}),
  \eprint{0903.1666}.

\bibitem[{\citenamefont{Lee}(2009)}]{Lee:2008fa}
\bibinfo{author}{\bibfnamefont{D.}~\bibnamefont{Lee}}, \bibinfo{journal}{Prog.
  Part. Nucl. Phys.} \textbf{\bibinfo{volume}{63}}, \bibinfo{pages}{117}
  (\bibinfo{year}{2009}), \eprint{arXiv:0804.3501 [nucl-th]}.

\bibitem[{\citenamefont{van Kolck}(1999)}]{vanKolck:1999mw}
\bibinfo{author}{\bibfnamefont{U.}~\bibnamefont{van Kolck}},
  \bibinfo{journal}{Prog. Part. Nucl. Phys.} \textbf{\bibinfo{volume}{43}},
  \bibinfo{pages}{337} (\bibinfo{year}{1999}), \eprint{nucl-th/9902015}.

\bibitem[{\citenamefont{Bedaque and van Kolck}(2002)}]{Bedaque:2002mn}
\bibinfo{author}{\bibfnamefont{P.~F.} \bibnamefont{Bedaque}} \bibnamefont{and}
  \bibinfo{author}{\bibfnamefont{U.}~\bibnamefont{van Kolck}},
  \bibinfo{journal}{Ann. Rev. Nucl. Part. Sci.} \textbf{\bibinfo{volume}{52}},
  \bibinfo{pages}{339} (\bibinfo{year}{2002}), \eprint{nucl-th/0203055}.

\bibitem[{\citenamefont{Epelbaum}(2006)}]{Epelbaum:2005pn}
\bibinfo{author}{\bibfnamefont{E.}~\bibnamefont{Epelbaum}},
  \bibinfo{journal}{Prog. Part. Nucl. Phys.} \textbf{\bibinfo{volume}{57}},
  \bibinfo{pages}{654} (\bibinfo{year}{2006}), \eprint{nucl-th/0509032}.

\bibitem[{\citenamefont{Epelbaum
  et~al.}(2009{\natexlab{c}})\citenamefont{Epelbaum, Hammer, and
  Mei{\ss}ner}}]{Epelbaum:2008ga}
\bibinfo{author}{\bibfnamefont{E.}~\bibnamefont{Epelbaum}},
  \bibinfo{author}{\bibfnamefont{H.-W.} \bibnamefont{Hammer}},
  \bibnamefont{and} \bibinfo{author}{\bibfnamefont{U.-G.}
  \bibnamefont{Mei{\ss}ner}}, \bibinfo{journal}{Rev. Mod. Phys.}
  \textbf{\bibinfo{volume}{81}}, \bibinfo{pages}{1773}
  (\bibinfo{year}{2009}{\natexlab{c}}), \eprint{arXiv:0811.1338 [nucl-th]}.

\bibitem[{\citenamefont{Epelbaum et~al.}(2010)\citenamefont{Epelbaum, Krebs,
  Lee, and Mei{\ss}ner}}]{Epelbaum:2009pd}
\bibinfo{author}{\bibfnamefont{E.}~\bibnamefont{Epelbaum}},
  \bibinfo{author}{\bibfnamefont{H.}~\bibnamefont{Krebs}},
  \bibinfo{author}{\bibfnamefont{D.}~\bibnamefont{Lee}}, \bibnamefont{and}
  \bibinfo{author}{\bibfnamefont{U.-G.} \bibnamefont{Mei{\ss}ner}},
  \bibinfo{journal}{Phys. Rev. Lett.} \textbf{\bibinfo{volume}{104}},
  \bibinfo{pages}{142501} (\bibinfo{year}{2010}), \eprint{0912.4195}.

\bibitem[{\citenamefont{Weinberg}(1990)}]{Weinberg:1990rz}
\bibinfo{author}{\bibfnamefont{S.}~\bibnamefont{Weinberg}},
  \bibinfo{journal}{Phys. Lett.} \textbf{\bibinfo{volume}{B251}},
  \bibinfo{pages}{288} (\bibinfo{year}{1990}).

\bibitem[{\citenamefont{Weinberg}(1991)}]{Weinberg:1991um}
\bibinfo{author}{\bibfnamefont{S.}~\bibnamefont{Weinberg}},
  \bibinfo{journal}{Nucl. Phys.} \textbf{\bibinfo{volume}{B363}},
  \bibinfo{pages}{3} (\bibinfo{year}{1991}).

\bibitem[{\citenamefont{Ordonez and van Kolck}(1992)}]{Ordonez:1992xp}
\bibinfo{author}{\bibfnamefont{C.}~\bibnamefont{Ordonez}} \bibnamefont{and}
  \bibinfo{author}{\bibfnamefont{U.}~\bibnamefont{van Kolck}},
  \bibinfo{journal}{Phys. Lett.} \textbf{\bibinfo{volume}{B291}},
  \bibinfo{pages}{459} (\bibinfo{year}{1992}).

\bibitem[{\citenamefont{Ordonez et~al.}(1994)\citenamefont{Ordonez, Ray, and
  van Kolck}}]{Ordonez:1993tn}
\bibinfo{author}{\bibfnamefont{C.}~\bibnamefont{Ordonez}},
  \bibinfo{author}{\bibfnamefont{L.}~\bibnamefont{Ray}}, \bibnamefont{and}
  \bibinfo{author}{\bibfnamefont{U.}~\bibnamefont{van Kolck}},
  \bibinfo{journal}{Phys. Rev. Lett.} \textbf{\bibinfo{volume}{72}},
  \bibinfo{pages}{1982} (\bibinfo{year}{1994}).

\bibitem[{\citenamefont{Ordonez et~al.}(1996)\citenamefont{Ordonez, Ray, and
  van Kolck}}]{Ordonez:1996rz}
\bibinfo{author}{\bibfnamefont{C.}~\bibnamefont{Ordonez}},
  \bibinfo{author}{\bibfnamefont{L.}~\bibnamefont{Ray}}, \bibnamefont{and}
  \bibinfo{author}{\bibfnamefont{U.}~\bibnamefont{van Kolck}},
  \bibinfo{journal}{Phys. Rev.} \textbf{\bibinfo{volume}{C53}},
  \bibinfo{pages}{2086} (\bibinfo{year}{1996}), \eprint{hep-ph/9511380}.

\bibitem[{\citenamefont{Epelbaum et~al.}(1998)\citenamefont{Epelbaum, Glockle,
  and Mei{\ss}ner}}]{Epelbaum:1998ka}
\bibinfo{author}{\bibfnamefont{E.}~\bibnamefont{Epelbaum}},
  \bibinfo{author}{\bibfnamefont{W.}~\bibnamefont{Glockle}}, \bibnamefont{and}
  \bibinfo{author}{\bibfnamefont{U.-G.} \bibnamefont{Mei{\ss}ner}},
  \bibinfo{journal}{Nucl. Phys.} \textbf{\bibinfo{volume}{A637}},
  \bibinfo{pages}{107} (\bibinfo{year}{1998}), \eprint{nucl-th/9801064}.

\bibitem[{\citenamefont{Epelbaum et~al.}(2000)\citenamefont{Epelbaum, Gloeckle,
  and Mei{\ss}ner}}]{Epelbaum:1999dj}
\bibinfo{author}{\bibfnamefont{E.}~\bibnamefont{Epelbaum}},
  \bibinfo{author}{\bibfnamefont{W.}~\bibnamefont{Gloeckle}}, \bibnamefont{and}
  \bibinfo{author}{\bibfnamefont{U.-G.} \bibnamefont{Mei{\ss}ner}},
  \bibinfo{journal}{Nucl. Phys.} \textbf{\bibinfo{volume}{A671}},
  \bibinfo{pages}{295} (\bibinfo{year}{2000}), \eprint{nucl-th/9910064}.

\bibitem[{\citenamefont{Friar and Coon}(1994)}]{Friar:1994}
\bibinfo{author}{\bibfnamefont{J.~L.} \bibnamefont{Friar}} \bibnamefont{and}
  \bibinfo{author}{\bibfnamefont{S.~A.} \bibnamefont{Coon}},
  \bibinfo{journal}{Phys. Rev.} \textbf{\bibinfo{volume}{C49}},
  \bibinfo{pages}{1272} (\bibinfo{year}{1994}).

\bibitem[{\citenamefont{Kaiser et~al.}(1997)\citenamefont{Kaiser, Brockmann,
  and Weise}}]{Kaiser:1997mw}
\bibinfo{author}{\bibfnamefont{N.}~\bibnamefont{Kaiser}},
  \bibinfo{author}{\bibfnamefont{R.}~\bibnamefont{Brockmann}},
  \bibnamefont{and} \bibinfo{author}{\bibfnamefont{W.}~\bibnamefont{Weise}},
  \bibinfo{journal}{Nucl. Phys.} \textbf{\bibinfo{volume}{A625}},
  \bibinfo{pages}{758} (\bibinfo{year}{1997}), \eprint{nucl-th/9706045}.

\bibitem[{\citenamefont{van Kolck}(1994)}]{vanKolck:1994yi}
\bibinfo{author}{\bibfnamefont{U.}~\bibnamefont{van Kolck}},
  \bibinfo{journal}{Phys. Rev.} \textbf{\bibinfo{volume}{C49}},
  \bibinfo{pages}{2932} (\bibinfo{year}{1994}).

\bibitem[{\citenamefont{Friar et~al.}(1999)\citenamefont{Friar, Huber, and van
  Kolck}}]{Friar:1998zt}
\bibinfo{author}{\bibfnamefont{J.~L.} \bibnamefont{Friar}},
  \bibinfo{author}{\bibfnamefont{D.}~\bibnamefont{Huber}}, \bibnamefont{and}
  \bibinfo{author}{\bibfnamefont{U.}~\bibnamefont{van Kolck}},
  \bibinfo{journal}{Phys. Rev.} \textbf{\bibinfo{volume}{C59}},
  \bibinfo{pages}{53} (\bibinfo{year}{1999}), \eprint{nucl-th/9809065}.

\bibitem[{\citenamefont{Epelbaum et~al.}(2002)\citenamefont{Epelbaum, Nogga,
  Gl{\"o}ckle, Kamada, Mei{\ss}ner, and Witala}}]{Epelbaum:2002vt}
\bibinfo{author}{\bibfnamefont{E.}~\bibnamefont{Epelbaum}},
  \bibinfo{author}{\bibfnamefont{A.}~\bibnamefont{Nogga}},
  \bibinfo{author}{\bibfnamefont{W.}~\bibnamefont{Gl{\"o}ckle}},
  \bibinfo{author}{\bibfnamefont{H.}~\bibnamefont{Kamada}},
  \bibinfo{author}{\bibfnamefont{U.-G.} \bibnamefont{Mei{\ss}ner}},
  \bibnamefont{and} \bibinfo{author}{\bibfnamefont{H.}~\bibnamefont{Witala}},
  \bibinfo{journal}{Phys. Rev.} \textbf{\bibinfo{volume}{C66}},
  \bibinfo{pages}{064001} (\bibinfo{year}{2002}), \eprint{nucl-th/0208023}.

\bibitem[{\citenamefont{Bernard et~al.}(1995)\citenamefont{Bernard, Kaiser, and
  Mei{\ss}ner}}]{Bernard:1995dp}
\bibinfo{author}{\bibfnamefont{V.}~\bibnamefont{Bernard}},
  \bibinfo{author}{\bibfnamefont{N.}~\bibnamefont{Kaiser}}, \bibnamefont{and}
  \bibinfo{author}{\bibfnamefont{U.-G.} \bibnamefont{Mei{\ss}ner}},
  \bibinfo{journal}{Int. J. Mod. Phys.} \textbf{\bibinfo{volume}{E4}},
  \bibinfo{pages}{193} (\bibinfo{year}{1995}), \eprint{hep-ph/9501384}.

\bibitem[{\citenamefont{B{\"u}ttiker and Mei{\ss}ner}(2000)}]{Buettiker:1999ap}
\bibinfo{author}{\bibfnamefont{P.}~\bibnamefont{B{\"u}ttiker}}
  \bibnamefont{and} \bibinfo{author}{\bibfnamefont{U.-G.}
  \bibnamefont{Mei{\ss}ner}}, \bibinfo{journal}{Nucl. Phys.}
  \textbf{\bibinfo{volume}{A668}}, \bibinfo{pages}{97} (\bibinfo{year}{2000}),
  \eprint{hep-ph/9908247}.

\bibitem[{\citenamefont{Bedaque et~al.}(2000)\citenamefont{Bedaque, Hammer, and
  van Kolck}}]{Bedaque:1999ve}
\bibinfo{author}{\bibfnamefont{P.~F.} \bibnamefont{Bedaque}},
  \bibinfo{author}{\bibfnamefont{H.-W.} \bibnamefont{Hammer}},
  \bibnamefont{and} \bibinfo{author}{\bibfnamefont{U.}~\bibnamefont{van
  Kolck}}, \bibinfo{journal}{Nucl. Phys.} \textbf{\bibinfo{volume}{A676}},
  \bibinfo{pages}{357} (\bibinfo{year}{2000}), \eprint{nucl-th/9906032}.

\bibitem[{\citenamefont{van Kolck et~al.}(1996)\citenamefont{van Kolck, Friar,
  and Goldman}}]{vanKolck:1996rm}
\bibinfo{author}{\bibfnamefont{U.}~\bibnamefont{van Kolck}},
  \bibinfo{author}{\bibfnamefont{J.~L.} \bibnamefont{Friar}}, \bibnamefont{and}
  \bibinfo{author}{\bibfnamefont{J.~T.} \bibnamefont{Goldman}},
  \bibinfo{journal}{Phys. Lett.} \textbf{\bibinfo{volume}{B371}},
  \bibinfo{pages}{169} (\bibinfo{year}{1996}), \eprint{nucl-th/9601009}.

\bibitem[{\citenamefont{van Kolck et~al.}(1998)\citenamefont{van Kolck,
  Rentmeester, Friar, Goldman, and de~Swart}}]{vanKolck:1997fu}
\bibinfo{author}{\bibfnamefont{U.}~\bibnamefont{van Kolck}},
  \bibinfo{author}{\bibfnamefont{M.~C.~M.} \bibnamefont{Rentmeester}},
  \bibinfo{author}{\bibfnamefont{J.~L.} \bibnamefont{Friar}},
  \bibinfo{author}{\bibfnamefont{J.~T.} \bibnamefont{Goldman}},
  \bibnamefont{and} \bibinfo{author}{\bibfnamefont{J.~J.}
  \bibnamefont{de~Swart}}, \bibinfo{journal}{Phys. Rev. Lett.}
  \textbf{\bibinfo{volume}{80}}, \bibinfo{pages}{4386} (\bibinfo{year}{1998}),
  \eprint{nucl-th/9710067}.

\bibitem[{\citenamefont{Epelbaum and Mei{\ss}ner}(1999)}]{Epelbaum:1999zn}
\bibinfo{author}{\bibfnamefont{E.}~\bibnamefont{Epelbaum}} \bibnamefont{and}
  \bibinfo{author}{\bibfnamefont{U.-G.} \bibnamefont{Mei{\ss}ner}},
  \bibinfo{journal}{Phys. Lett.} \textbf{\bibinfo{volume}{B461}},
  \bibinfo{pages}{287} (\bibinfo{year}{1999}), \eprint{nucl-th/9902042}.

\bibitem[{\citenamefont{Friar and van Kolck}(1999)}]{Friar:1999zr}
\bibinfo{author}{\bibfnamefont{J.~L.} \bibnamefont{Friar}} \bibnamefont{and}
  \bibinfo{author}{\bibfnamefont{U.}~\bibnamefont{van Kolck}},
  \bibinfo{journal}{Phys. Rev.} \textbf{\bibinfo{volume}{C60}},
  \bibinfo{pages}{034006} (\bibinfo{year}{1999}), \eprint{nucl-th/9906048}.

\bibitem[{\citenamefont{Walzl et~al.}(2001)\citenamefont{Walzl, Mei{\ss}ner,
  and Epelbaum}}]{Walzl:2000cx}
\bibinfo{author}{\bibfnamefont{M.}~\bibnamefont{Walzl}},
  \bibinfo{author}{\bibfnamefont{U.~G.} \bibnamefont{Mei{\ss}ner}},
  \bibnamefont{and} \bibinfo{author}{\bibfnamefont{E.}~\bibnamefont{Epelbaum}},
  \bibinfo{journal}{Nucl. Phys.} \textbf{\bibinfo{volume}{A693}},
  \bibinfo{pages}{663} (\bibinfo{year}{2001}), \eprint{nucl-th/0010019}.

\bibitem[{\citenamefont{Friar et~al.}(2003)\citenamefont{Friar, van Kolck,
  Payne, and Coon}}]{Friar:2003yv}
\bibinfo{author}{\bibfnamefont{J.~L.} \bibnamefont{Friar}},
  \bibinfo{author}{\bibfnamefont{U.}~\bibnamefont{van Kolck}},
  \bibinfo{author}{\bibfnamefont{G.~L.} \bibnamefont{Payne}}, \bibnamefont{and}
  \bibinfo{author}{\bibfnamefont{S.~A.} \bibnamefont{Coon}},
  \bibinfo{journal}{Phys. Rev.} \textbf{\bibinfo{volume}{C68}},
  \bibinfo{pages}{024003} (\bibinfo{year}{2003}), \eprint{nucl-th/0303058}.

\bibitem[{\citenamefont{Epelbaum et~al.}(2005)\citenamefont{Epelbaum,
  Mei{\ss}ner, and Palomar}}]{Epelbaum:2004xf}
\bibinfo{author}{\bibfnamefont{E.}~\bibnamefont{Epelbaum}},
  \bibinfo{author}{\bibfnamefont{U.-G.} \bibnamefont{Mei{\ss}ner}},
  \bibnamefont{and} \bibinfo{author}{\bibfnamefont{J.~E.}
  \bibnamefont{Palomar}}, \bibinfo{journal}{Phys. Rev.}
  \textbf{\bibinfo{volume}{C71}}, \bibinfo{pages}{024001}
  (\bibinfo{year}{2005}), \eprint{nucl-th/0407037}.

\bibitem[{\citenamefont{Epelbaum and Mei{\ss}ner}(2005)}]{Epelbaum:2005fd}
\bibinfo{author}{\bibfnamefont{E.}~\bibnamefont{Epelbaum}} \bibnamefont{and}
  \bibinfo{author}{\bibfnamefont{U.-G.} \bibnamefont{Mei{\ss}ner}},
  \bibinfo{journal}{Phys. Rev.} \textbf{\bibinfo{volume}{C72}},
  \bibinfo{pages}{044001} (\bibinfo{year}{2005}), \eprint{nucl-th/0502052}.

\bibitem[{\citenamefont{Borasoy
  et~al.}(2007{\natexlab{b}})\citenamefont{Borasoy, Epelbaum, Krebs, Lee, and
  Mei{\ss}ner}}]{Borasoy:2007vy}
\bibinfo{author}{\bibfnamefont{B.}~\bibnamefont{Borasoy}},
  \bibinfo{author}{\bibfnamefont{E.}~\bibnamefont{Epelbaum}},
  \bibinfo{author}{\bibfnamefont{H.}~\bibnamefont{Krebs}},
  \bibinfo{author}{\bibfnamefont{D.}~\bibnamefont{Lee}}, \bibnamefont{and}
  \bibinfo{author}{\bibfnamefont{U.-G.} \bibnamefont{Mei{\ss}ner}},
  \bibinfo{journal}{Eur. Phys. J.} \textbf{\bibinfo{volume}{A34}},
  \bibinfo{pages}{185} (\bibinfo{year}{2007}{\natexlab{b}}),
  \eprint{arXiv:0708.1780 [nucl-th]}.

\bibitem[{\citenamefont{Stoks et~al.}(1993)\citenamefont{Stoks, Kompl,
  Rentmeester, and de~Swart}}]{Stoks:1993tb}
\bibinfo{author}{\bibfnamefont{V.~G.~J.} \bibnamefont{Stoks}},
  \bibinfo{author}{\bibfnamefont{R.~A.~M.} \bibnamefont{Kompl}},
  \bibinfo{author}{\bibfnamefont{M.~C.~M.} \bibnamefont{Rentmeester}},
  \bibnamefont{and} \bibinfo{author}{\bibfnamefont{J.~J.}
  \bibnamefont{de~Swart}}, \bibinfo{journal}{Phys. Rev.}
  \textbf{\bibinfo{volume}{C48}}, \bibinfo{pages}{792} (\bibinfo{year}{1993}).

\bibitem[{\citenamefont{Gonzalez~Trotter
  et~al.}(1999)}]{GonzalezTrotter:1999zt}
\bibinfo{author}{\bibfnamefont{D.~E.} \bibnamefont{Gonzalez~Trotter}}
  \bibnamefont{et~al.}, \bibinfo{journal}{Phys. Rev. Lett.}
  \textbf{\bibinfo{volume}{83}}, \bibinfo{pages}{3788} (\bibinfo{year}{1999}),
  \eprint{nucl-ex/9904011}.

\bibitem[{\citenamefont{Huhn et~al.}(2000)\citenamefont{Huhn, W\"atzold, Weber,
  Siepe, von Witsch, Wita\l{}a, and Gl\"ockle}}]{Huhn:2000a}
\bibinfo{author}{\bibfnamefont{V.}~\bibnamefont{Huhn}},
  \bibinfo{author}{\bibfnamefont{L.}~\bibnamefont{W\"atzold}},
  \bibinfo{author}{\bibfnamefont{C.}~\bibnamefont{Weber}},
  \bibinfo{author}{\bibfnamefont{A.}~\bibnamefont{Siepe}},
  \bibinfo{author}{\bibfnamefont{W.}~\bibnamefont{von Witsch}},
  \bibinfo{author}{\bibfnamefont{H.}~\bibnamefont{Wita\l{}a}},
  \bibnamefont{and}
  \bibinfo{author}{\bibfnamefont{W.}~\bibnamefont{Gl\"ockle}},
  \bibinfo{journal}{Phys. Rev. Lett.} \textbf{\bibinfo{volume}{85}},
  \bibinfo{pages}{1190} (\bibinfo{year}{2000}).

\bibitem[{\citenamefont{Gonzalez~Trotter
  et~al.}(2006)\citenamefont{Gonzalez~Trotter, Meneses, Tornow, Howell, Chen,
  Crowell, Roper, Walter, Schmidt, Wita\l{}a et~al.}}]{GonzalezTrotter:2006a}
\bibinfo{author}{\bibfnamefont{D.~E.} \bibnamefont{Gonzalez~Trotter}},
  \bibinfo{author}{\bibfnamefont{F.~S.} \bibnamefont{Meneses}},
  \bibinfo{author}{\bibfnamefont{W.}~\bibnamefont{Tornow}},
  \bibinfo{author}{\bibfnamefont{C.~R.} \bibnamefont{Howell}},
  \bibinfo{author}{\bibfnamefont{Q.}~\bibnamefont{Chen}},
  \bibinfo{author}{\bibfnamefont{A.~S.} \bibnamefont{Crowell}},
  \bibinfo{author}{\bibfnamefont{C.~D.} \bibnamefont{Roper}},
  \bibinfo{author}{\bibfnamefont{R.~L.} \bibnamefont{Walter}},
  \bibinfo{author}{\bibfnamefont{D.}~\bibnamefont{Schmidt}},
  \bibinfo{author}{\bibfnamefont{H.}~\bibnamefont{Wita\l{}a}},
  \bibnamefont{et~al.}, \bibinfo{journal}{Phys. Rev. C}
  \textbf{\bibinfo{volume}{73}}, \bibinfo{pages}{034001}
  (\bibinfo{year}{2006}).

\bibitem[{\citenamefont{Chen et~al.}(2008)\citenamefont{Chen, Howell, Carman,
  Gibbs, Gibson, Hussein, Kiser, Mertens, Moore, Morris et~al.}}]{Chen:2008a}
\bibinfo{author}{\bibfnamefont{Q.}~\bibnamefont{Chen}},
  \bibinfo{author}{\bibfnamefont{C.~R.} \bibnamefont{Howell}},
  \bibinfo{author}{\bibfnamefont{T.~S.} \bibnamefont{Carman}},
  \bibinfo{author}{\bibfnamefont{W.~R.} \bibnamefont{Gibbs}},
  \bibinfo{author}{\bibfnamefont{B.~F.} \bibnamefont{Gibson}},
  \bibinfo{author}{\bibfnamefont{A.}~\bibnamefont{Hussein}},
  \bibinfo{author}{\bibfnamefont{M.~R.} \bibnamefont{Kiser}},
  \bibinfo{author}{\bibfnamefont{G.}~\bibnamefont{Mertens}},
  \bibinfo{author}{\bibfnamefont{C.~F.} \bibnamefont{Moore}},
  \bibinfo{author}{\bibfnamefont{C.}~\bibnamefont{Morris}},
  \bibnamefont{et~al.}, \bibinfo{journal}{Phys. Rev. C}
  \textbf{\bibinfo{volume}{77}}, \bibinfo{pages}{054002}
  (\bibinfo{year}{2008}).

\bibitem[{\citenamefont{Stapp et~al.}(1957)\citenamefont{Stapp, Ypsilantis, and
  Metropolis}}]{Stapp:1956mz}
\bibinfo{author}{\bibfnamefont{H.~P.} \bibnamefont{Stapp}},
  \bibinfo{author}{\bibfnamefont{T.~J.} \bibnamefont{Ypsilantis}},
  \bibnamefont{and}
  \bibinfo{author}{\bibfnamefont{N.}~\bibnamefont{Metropolis}},
  \bibinfo{journal}{Phys. Rev.} \textbf{\bibinfo{volume}{105}},
  \bibinfo{pages}{302} (\bibinfo{year}{1957}).

\bibitem[{\citenamefont{L{\"u}scher}(1986)}]{Luscher:1985dn}
\bibinfo{author}{\bibfnamefont{M.}~\bibnamefont{L{\"u}scher}},
  \bibinfo{journal}{Commun. Math. Phys.} \textbf{\bibinfo{volume}{104}},
  \bibinfo{pages}{177} (\bibinfo{year}{1986}).

\bibitem[{\citenamefont{Gazit et~al.}(2009)\citenamefont{Gazit, Quaglioni, and
  Navratil}}]{Gazit:2008ma}
\bibinfo{author}{\bibfnamefont{D.}~\bibnamefont{Gazit}},
  \bibinfo{author}{\bibfnamefont{S.}~\bibnamefont{Quaglioni}},
  \bibnamefont{and} \bibinfo{author}{\bibfnamefont{P.}~\bibnamefont{Navratil}},
  \bibinfo{journal}{Phys. Rev. Lett.} \textbf{\bibinfo{volume}{103}},
  \bibinfo{pages}{102502} (\bibinfo{year}{2009}), \eprint{0812.4444}.

\bibitem[{\citenamefont{Lee}(2006{\natexlab{a}})}]{Lee:2005xy}
\bibinfo{author}{\bibfnamefont{D.}~\bibnamefont{Lee}}, \bibinfo{journal}{Phys.
  Rev.} \textbf{\bibinfo{volume}{A73}}, \bibinfo{pages}{063204}
  (\bibinfo{year}{2006}{\natexlab{a}}), \eprint{physics/0512085}.

\bibitem[{\citenamefont{Wigner}(1937)}]{Wigner:1937}
\bibinfo{author}{\bibfnamefont{E.}~\bibnamefont{Wigner}},
  \bibinfo{journal}{Phys. Rev.} \textbf{\bibinfo{volume}{51}},
  \bibinfo{pages}{106} (\bibinfo{year}{1937}).

\bibitem[{\citenamefont{Lee}(2005)}]{Lee:2004hc}
\bibinfo{author}{\bibfnamefont{D.}~\bibnamefont{Lee}}, \bibinfo{journal}{Phys.
  Rev.} \textbf{\bibinfo{volume}{C71}}, \bibinfo{pages}{044001}
  (\bibinfo{year}{2005}), \eprint{nucl-th/0407101}.

\bibitem[{\citenamefont{Chen et~al.}(2004)\citenamefont{Chen, Lee, and
  Sch{\"a}fer}}]{Chen:2004rq}
\bibinfo{author}{\bibfnamefont{J.-W.} \bibnamefont{Chen}},
  \bibinfo{author}{\bibfnamefont{D.}~\bibnamefont{Lee}}, \bibnamefont{and}
  \bibinfo{author}{\bibfnamefont{T.}~\bibnamefont{Sch{\"a}fer}},
  \bibinfo{journal}{Phys. Rev. Lett.} \textbf{\bibinfo{volume}{93}},
  \bibinfo{pages}{242302} (\bibinfo{year}{2004}), \eprint{nucl-th/0408043}.

\bibitem[{\citenamefont{Lee}(2007{\natexlab{a}})}]{Lee:2007eu}
\bibinfo{author}{\bibfnamefont{D.}~\bibnamefont{Lee}}, \bibinfo{journal}{Phys.
  Rev. Lett.} \textbf{\bibinfo{volume}{98}}, \bibinfo{pages}{182501}
  (\bibinfo{year}{2007}{\natexlab{a}}), \eprint{nucl-th/0701041}.

\bibitem[{\citenamefont{Lee}(2006{\natexlab{b}})}]{Lee:2005fk}
\bibinfo{author}{\bibfnamefont{D.}~\bibnamefont{Lee}}, \bibinfo{journal}{Phys.
  Rev.} \textbf{\bibinfo{volume}{B73}}, \bibinfo{pages}{115112}
  (\bibinfo{year}{2006}{\natexlab{b}}), \eprint{cond-mat/0511332}.

\bibitem[{\citenamefont{Lee}(2007{\natexlab{b}})}]{Lee:2006hr}
\bibinfo{author}{\bibfnamefont{D.}~\bibnamefont{Lee}}, \bibinfo{journal}{Phys.
  Rev.} \textbf{\bibinfo{volume}{B75}}, \bibinfo{pages}{134502}
  (\bibinfo{year}{2007}{\natexlab{b}}), \eprint{cond-mat/0606706}.

\bibitem[{\citenamefont{Hagen et~al.}(2007)\citenamefont{Hagen, Dean,
  Hjorth-Jensen, Papenbrock, and Schwenk}}]{Hagen:2007hi}
\bibinfo{author}{\bibfnamefont{G.}~\bibnamefont{Hagen}},
  \bibinfo{author}{\bibfnamefont{D.~J.} \bibnamefont{Dean}},
  \bibinfo{author}{\bibfnamefont{M.}~\bibnamefont{Hjorth-Jensen}},
  \bibinfo{author}{\bibfnamefont{T.}~\bibnamefont{Papenbrock}},
  \bibnamefont{and} \bibinfo{author}{\bibfnamefont{A.}~\bibnamefont{Schwenk}},
  \bibinfo{journal}{Phys. Rev.} \textbf{\bibinfo{volume}{C76}},
  \bibinfo{pages}{044305} (\bibinfo{year}{2007}), \eprint{0707.1516}.

\bibitem[{\citenamefont{Pieper}(2009)}]{Pieper:2009aa}
\bibinfo{author}{\bibfnamefont{S.~C.} \bibnamefont{Pieper}},
  \bibinfo{journal}{B. Am. Phys. Soc.} \textbf{\bibinfo{volume}{54}},
  \bibinfo{pages}{70} (\bibinfo{year}{2009}).

\bibitem[{\citenamefont{Maris et~al.}(2009)\citenamefont{Maris, Vary, and
  Shirokov}}]{Maris:2008ax}
\bibinfo{author}{\bibfnamefont{P.}~\bibnamefont{Maris}},
  \bibinfo{author}{\bibfnamefont{J.~P.} \bibnamefont{Vary}}, \bibnamefont{and}
  \bibinfo{author}{\bibfnamefont{A.~M.} \bibnamefont{Shirokov}},
  \bibinfo{journal}{Phys. Rev.} \textbf{\bibinfo{volume}{C79}},
  \bibinfo{pages}{014308} (\bibinfo{year}{2009}), \eprint{0808.3420}.

\bibitem[{\citenamefont{de~Forcrand and Fromm}(2010)}]{deForcrand:2009dh}
\bibinfo{author}{\bibfnamefont{P.}~\bibnamefont{de~Forcrand}} \bibnamefont{and}
  \bibinfo{author}{\bibfnamefont{M.}~\bibnamefont{Fromm}},
  \bibinfo{journal}{Phys. Rev. Lett.} \textbf{\bibinfo{volume}{104}},
  \bibinfo{pages}{112005} (\bibinfo{year}{2010}), \eprint{0907.1915}.

\end{thebibliography}

\end{document}